\documentclass[fleqn,usenatbib]{mnras}

\usepackage{newtxtext,newtxmath}

\usepackage[T1]{fontenc}
\usepackage{ae,aecompl}

\usepackage{graphicx}    
\usepackage{amsmath}    
\usepackage{amssymb}    
\usepackage{amssymb}    
\usepackage{amsbsy}    
\usepackage{xcolor}

\usepackage{float}
\title[Global properties of coupled stellar models]{On the impact of the structural surface effect on global stellar properties and asteroseismic analyses}

\author[A. C. S. J{\o}rgensen et al.]{%
Andreas Christ S{\o}lvsten J{\o}rgensen$^{1,2}$\thanks{E-mail: a.c.s.joergensen@bham.ac.uk},
Josefina Montalb{\'a}n$^{1}$,
George C. Angelou$^{2}$,
\and
Andrea Miglio$^{1}$,
Achim Weiss$^{2}$,
Richard Scuflaire$^{3}$,
Arlette Noels$^{3}$,
\and
Jakob R{\o}rsted Mosumgaard$^{4}$, and V\'ictor Silva Aguirre$^{4}$ \\
$^{1}$School of Physics and Astronomy, University of Birmingham, Edgbaston B15 2TT, UK \\
$^{2}$Max-Planck-Institut f\"ur Astrophysik, Karl-Schwarzschild-Str. 1, D-85748 Garching, Germany 
\\
$^{3}$Space sciences, Technologies and Astrophysics Research (STAR) Institute, Universit{\'e} de Li{\`e}ge, \\19C All{\'e}e du 6 Ao{\^u}t,B-4000 Li{\`e}ge, Belgium
\\
$^{4}$Stellar Astrophysics Centre (SAC), Department of Physics and Astronomy, Aarhus University, Ny Munkegade 120,\\ DK-8000 Aarhus C, Denmark
}

\date{Accepted XXX. Received YYY; in original form 23.06.2020}

\pubyear{2020}

\begin{document}
\label{firstpage}
\pagerange{\pageref{firstpage}--\pageref{lastpage}}
\maketitle
 
\begin{abstract}
In a series of papers, we have recently demonstrated that it is possible to construct stellar structure models that robustly mimic the stratification of multi-dimensional radiative magneto-hydrodynamic simulations at every time-step of the computed evolution. The resulting models offer a more realistic depiction of the near-surface layers of stars with convective envelopes than parameterizations, such as mixing length theory, do. In this paper, we explore how this model improvement impacts on seismic and non-seismic properties of stellar models across the Hertzsprung-Russell diagram. We show that the improved description of the outer boundary layers alters the predicted global stellar properties at different evolutionary stages. In a hare and hound exercise, we show that this plays a key role for asteroseismic analyses, as it, for instance, often shifts the inferred stellar age estimates by more than 10 per cent. Improper boundary conditions may thus introduce systematic errors that exceed the required accuracy of the PLATO space mission. {\color{black}Moreover}, we discuss different approximations for how to compute stellar oscillation frequencies. We demonstrate that the so-called gas $\Gamma_1$ approximation performs reasonably well for all main-sequence stars. {\color{black}Using} a Monte Carlo approach, we show that the model frequencies of our hybrid solar models are consistent with observations within the uncertainties of the global solar parameters when using the so-called reduced $\Gamma_1$ approximation.
\end{abstract}




\section{Introduction}

In asteroseismic analyses, stellar parameters, as well as the internal physical processes, are determined by comparing observations with theoretical stellar models. To give a holistic depiction of the entire structure and evolution of stars, current stellar models are subject to a set of simplifying assumptions. Stellar models thus assume spherical symmetry, which allows structures to be computed as a function of a single spatial coordinate. They are one-dimensional (1D). Furthermore, to capture the complicated behaviour of multi-dimensional physical processes such as turbulent convection, simplified parameterizations are employed. This includes mixing length theory \citep[MLT][]{Bohm-Vitense1958} and full-spectrum theory \citep[FST][]{Canuto1991,Canuto1992,Canuto1996}. Without these 1D parameterizations, it becomes intractable to compute the details of inherently dynamical processes over the nuclear timescale. However, the invoked simplifying assumptions do not perfectly capture the behaviours of the relevant hydrodynamic processes. In the case of superadiabatic convection, the resulting inadequate treatment of the surface layers of stars with convective envelopes is known to lead to a systematic offset between observations and the predicted model frequencies. This tension with data, i.e. the aforementioned frequency shift, is the so-called \textit{structural} surface effect.

In addition, model frequencies are computed under the assumption of adiabaticity. The neglect of non-adiabatic energetic and the contributions of turbulent pressure leads to yet another frequency offset known as the \textit{modal} surface effect. The combined structural and modal surface effect has haunted astero- and helioseismology for decades  \citep{Brown1984, jcd1989, 1990LNP...367..283G, Aerts2020}.

It is common practice to deal with the surface effect in the post-processing, using semi-empirical correction relations \citep[e.g.][]{Kjeldsen2008,Sonoi2015,Ball2014}. However, the versatility and broad applicability of these correction relations throughout the Hertzprung-Russell (HR) diagram is yet to be fully mapped. Indeed, several studies show that the use of different surface correction relations introduces systematic errors in the inferred stellar parameters from asteroseismic analyses \citep{Nsamba2018,Joergensen2019,Joergensen2020}. Even if this was not the case, the improper depiction of the boundary layers, from which the surface effect arises, would still introduce systematic offsets in the inferred stellar properties. This is because the surface effect, i.e. the frequency offset, is not the only consequence of an inadequate treatment of superadiabatic convection. Indeed, the improper depiction of the boundary layers has repeatedly been shown to affect the predicted stellar evolution tracks \citep{Salaris2015,Mosumgaard2017,Mosumgaard2018,Sonoi2019}.

Multi-dimensional simulations of radiative magneto-hydrodynamics (RHD) \citep[cf.][]{2012JCoPh.231..919F, Magic2013,Trampedach2013} yield a physically more realistic depiction of convection than stellar structure models do. 
However, such simulations cannot provide the same holistic depiction of stars as stellar models, due to their high computational cost. To overcome this issue, one might combine the advantages of both approaches by implementing the results from the physically more realistic multi-dimensional simulations into the holistic stellar models from stellar evolution codes. One way to do this is referred to as patching. 
In this procedure, the outermost layers of a given 1D stellar model are replaced by the average stratification of a multi-dimensional, often three-dimensional (3D), simulation \citep{Rosenthal1999,Piau2014,Sonoi2015, Ball2016, Magic2016, Joergensen2017,Trampedach2017,Manchon2018,Joergensen2019,Houdek2019}. Following the terminology introduced by \cite{Joergensen2018}, we will refer to such mean stratifications of the outer superadiabatic layers as $\langle 3 \mathrm{D }\rangle$-envelopes. We note that the employed $\langle 3 \mathrm{D }\rangle$-envelopes do by no means cover the entire convective zone. Indeed, they only reach down into the nearly-adiabatic region and are, therefore, often referred to as "3D-atmospheres" by other authors.

Due to a high degree of homology between the multi-dimensional simulations, it is possible to robustly recover the required $\langle 3 \mathrm{D }\rangle$-envelopes by means of interpolation \citep{Joergensen2017,Joergensen2019}. Patched models can thus be constructed across the HR diagram for any combination of effective temperature ($T_\mathrm{eff}$), surface gravity ($\log g$), and metallicity ($\mathrm{[Fe/H]}$).

Patched models do not suffer from the same structural deficiencies as standard stellar models and have repeatedly been shown to overcome the associated contributions to the surface effect \citep[e.g.][]{Rosenthal1999}. The remaining discrepancies between the predicted model frequencies and observations are modal, i.e. the remaining surface effect does not indicate shortcomings of the stellar structure models themselves.

While patching solves some of the structural inadequacies of 1D stellar models, patching only addresses the inadequacies of the model at the last time-step. Throughout the computed stellar evolution, the interior model has thus been subject to incorrect boundary conditions through the simplified assumptions that entered the surface layers. To overcome this issue, \cite{Joergensen2018} proposed a method for appending $\langle 3 \mathrm{D }\rangle$-envelopes \textit{at every time-step} and adjusting the interior model accordingly, using the $\langle 3 \mathrm{D }\rangle$-envelopes as outer boundary conditions. Using the terminology from \cite{Joergensen2018}, we refer to the implementation of the $\langle 3 \mathrm{D }\rangle$-envelopes into the stellar evolution code as the \textit{coupling} of 1D and 3D models. The resulting hybrid models are thus referred to as coupled models.

In a series of papers, we have explored the properties of coupled models. We have shown that the outermost layers of coupled models perfectly mimic the underlying 3D simulation \citep{Joergensen2018,Joergensen2019}. Furthermore, we have shown that the structures of coupled models are continuous in several physical quantities at the transition between the interior and the appended $\langle 3 \mathrm{D }\rangle$-envelope \citep{JoergensenAngelou2019}. We have demonstrated that coupled models mend the surface effect for the present-day Sun and overcome degeneracies of MLT \citep[cf.][]{JoergensenAngelou2019}. Finally, we have shown that the use of coupled models has significant consequences for stellar evolution tracks \citep[cf.][]{Mosumgaard2020}.

In this paper, we continue our exploration of the properties of coupled models, quantifying the implications of the improved boundary conditions across the HR diagram, and demonstrating the general efficacy of our methodology. 

The aim of the paper is thus threefold: first, we revisit the case of the present-day Sun (cf. Section~\ref{sec:solar}). By employing a Monte Carlo analysis, we quantify the uncertainties that are associated with the model frequencies of coupled models. We hereby aim to contribute to the discussion on whether current hybrid models, including coupled and patched models, perform to the level of precision of the asteroseismic data. 

Secondly, most authors, including ourselves, compute the model frequencies of hybrid stellar models, using the so-called gas $\Gamma_1$ approximation to avoid the complications that arise from computing model frequencies using adiabatic pulsation codes. However, there is no justification for this approach beyond the fact that it yields reasonable results for the present-day Sun. Whether this approach is generally valid across the HR diagram is hitherto unknown. We will address this issue in Section~\ref{sec:gasG1}, showing that the gas $\Gamma_1$ approximation does, indeed, perform equally well for other {\color{black}low-mass main-sequence} stars. 

Finally, having discussed the accuracy and proven the versatility of our coupling scheme, we quantify the implications of improving the outer boundary conditions across the HR diagram (cf. Section~\ref{sec:global} and \ref{sec:hare}). Here, we address both seismic and non-seismic stellar parameters and properties, including the stellar ages.


\section{Coupled stellar models} \label{sec:method}

{\color{black}Standard stellar structure models commonly use semi-empirical or theoretical relations between the temperature ($T$) and the optical depth ($\tau$) to depict the atmospheric stratification above the photosphere. Such $T(\tau)$ relations set the outer boundary conditions for the interior structure \citep[e.g.][]{Weiss2008, Kippenhahn}. They include Eddington grey atmospheres or the semi-empirical relations by \cite{KrishnaSwamy1966} and \cite{1981ApJS...45..635V}.

Our coupled stellar models, on the other hand, draw upon $\langle \mathrm{3D} \rangle$-envelopes to set the outer boundary conditions and to depict the outermost layers. We stress that this is the case \textit{at every time-step of the evolution}. The stratification of the $\langle \mathrm{3D} \rangle$-envelopes are determined by interpolation in an existing grid of 3D simulations at every iteration. For this purpose, we use the interpolation scheme by \cite{Joergensen2017} and \cite{Joergensen2019}. This method robustly recovers the accurate mean stratification of the underlying 3D simulations by interpolating in the effective temperature ($T_\mathrm{eff}$), surface gravity ($g$), and metallicity ($\mathrm{[Fe/H]}$). While the low number of available 3D simulations have introduced interpolation errors on the red giant branch (RGB) in previous papers, this issue has now been overcome as demonstrated in Appendix~\ref{sec:interr}.}

In contrast to $T(\tau)$ relations, the $\langle \mathrm{3D} \rangle$-envelopes stretch into the nearly-adiabatic region of the convective zone, placing the outer boundary condition far below the photosphere. Throughout the paper, we set the base of the envelope at a thermal pressure that is 16 times larger than the pressure at the density inflexion at the stellar surface --- the same criterion was used in previous papers \citep{Joergensen2018, Joergensen2019, JoergensenAngelou2019, Mosumgaard2020}. We thus define the point, at which we supply the outer boundary conditions, based on the pressure. {\color{black}This} implies that the physical extent of the appended envelope varies from model to model. 
For the present-day Sun, the outer boundary conditions are placed more than one thousand kilometres below the surface.

{\color{black}By construction, the temperature and thermal pressure stratification of the resulting coupled models are continuous at the transition between the interior structure and the appended $\langle \mathrm{3D} \rangle$-envelope. All quantities that are derived from the equation of state (EOS) and the opacity tables are, therefore, likewise continuous. Moreover, the implementation ensures that the Stefan-Boltzmann law is fulfilled. Finally, the employed input physic is chosen in such a way as to achieve a high level of consistency between the coupled models and the underlying 3D simulations. For instance, throughout this paper, we use the composition found by \cite{Asplund2009} (AGSS09). We refer to \cite{Joergensen2018} and \cite{JoergensenWeiss2019} for further details on our coupling scheme (cf. the flowchart in Fig.~1 of \citealt{JoergensenWeiss2019}).}

In this paper, we compute coupled stellar models using the Garching Stellar Evolution Code \citep[\textsc{garstec}][]{Weiss2008} and the \textsc{cl{\'e}s} \citep[Code Li{\'e}geois d'{\'Evolution Stellaire;}][]{CLES} stellar evolution code. We hereby show that the presented results are supported by independent stellar evolution codes. In all cases, we draw upon the Stagger-grid 3D RHD simulations by \cite{Magic2013}. Coupled models were computed for the first time by \cite{Joergensen2018} using \textsc{garstec}. Indeed, results presented on coupled models in previous papers were all computed using \textsc{garstec}, making results from this code an important reference. We have now included the same procedures into the \textsc{cl{\'e}s} stellar evolution code, and we mainly perform computations using \textsc{cl{\'e}s} in this paper.

We compute model frequencies for stellar pulsations, using the Aarhus adiabatic pulsation package, \textsc{adipls} \citep{Christensen-Dalsgaard2008a}. Due to the inclusion of turbulent pressure, we compute the stellar oscillation frequencies within the so-called reduced and gas $\Gamma_1$ approximations. For a thorough introduction to both approximations, we refer the reader to \cite{Rosenthal1999} and \cite{Houdek2017}. With the exception of Section~\ref{sec:solar}, we deploy the gas $\Gamma_1$ approximation throughout this paper. {\color{black}While both the reduced and gas $\Gamma_1$ approximations are the state of the art and widely used \citep[e.g.][]{Sonoi2015}, we note that that the underlying assumptions on how to treat turbulent pressure in adiabatic oscillation codes have only been tested in a limited number of cases \cite[e.g.][]{Houdek2017}. We, therefore, explore the validity of the gas $\Gamma_1$ approximation in Section~\ref{sec:gasG1}. The use of a fully non-adiabatic time-dependent stellar oscillation code that would overcome the limitations of the reduced and gas $\Gamma_1$ approximations is beyond the scope of this paper.}

For all presented models, we draw upon MLT. In standard stellar models, the associated mixing length parameter ($\alpha_\textsc{mlt}$) must bridge the entropy difference between the deep adiabat and the photosphere. When dealing with coupled stellar models, on the other hand, the appended $\langle \mathrm{3D} \rangle$-envelopes covers most of the superadiabatic region, stretching far below the photosphere. However, we still need MLT to bridge the entropy jump between the base of the $\langle \mathrm{3D} \rangle$-envelope and the deep adiabat. In coupled stellar models, MLT is thus used to describe a narrow nearly-adiabatic layer. As a result, $\alpha_\textsc{mlt}$ plays a different role in coupled stellar models than in standard stellar models, encompassing very different information in the two scenarios. When dealing with coupled stellar models, solar calibrations with different input physics might thus yield significantly different values for $\alpha_\textsc{mlt}$, and these values might by far exceeds the values encountered for standard stellar models. For a more detailed discussion on this issue, we refer the reader to \cite{JoergensenAngelou2019} and \cite{Mosumgaard2020}.


\section{The present-day Sun} \label{sec:thesun}

While solar calibrations involving coupled models are already to be found in the literature \citep[e.g.][]{JoergensenWeiss2019}, the uncertainties on the obtained stellar properties have not yet been quantified, making a direct interpretation less tangible. By performing an MCMC analysis, we address this issue by mapping the uncertainties on the derived stellar properties including the individual stellar oscillation frequencies. We do so within both the gas and the reduced $\Gamma_1$ approximations. Uncertainties for standard stellar models have been quantified by, e.g. \cite{Bahcall2006}, \cite{2010ApJ...719..865S}, \cite{2013PhRvD..87d3001S}, \cite{Vinyoles2017}, and \cite{2020arXiv200406365V}.


\subsection{MCMC {algorithms}} \label{sec:MCMC}

Monte Carlo methods have proven to be exceedingly fruitful techniques for Bayesian inference and are employed within many fields of astrophysics \citep[e.g.][]{Bahcall2006, Handberg2011, Bazot2012, Lund2017, Vinyoles2017, 2019ApJ...887L...1B, Porqueres2019, Porqueres2019Lya}. Much can be learned from these studies since they give a thorough mapping of posterior probability distributions rather than solely providing a best-fitting model.

In this paper, we use the algorithm \textsc{hephaestus} described by \cite{JoergensenAngelou2019} to perform the study presented in Section~\ref{sec:solar2}. \textsc{hephaestus} is a stellar model optimisation and search pipeline that employs an MCMC algorithm based on the MCMC ensemble sampler published by \cite{emcee}. The underlying procedure for this ensemble sampler was originally designed by \cite{Goodman2010}.

In short, \textsc{hephaestus} engages several walkers that map the space spanned by the selected global parameters of stellar models. In this process, each walker constructs a Markov chain. For each entry in a Markov chain, the associated walker computes the evolution of a star up until a certain age using \textsc{garstec}. 
The global parameters of each of the models, including the stellar age, are randomly drawn from proposal distributions around the parameters of the previous samples in the Markov chains of a subset of the other walkers. By comparing seismic and and non-seismic properties of the final structure model from the computed evolution track to observations, \textsc{hephaestus} evaluates the posterior probability of the constructed model --- we specify the likelihood in Section~\ref{sec:solar}. Based on this comparison, \textsc{hephaestus} either rejects or accepts the investigated models as an entry in the Markov chain. Following this procedure, the density of the accumulated samples across the parameter space converge towards the posterior probability distribution of the stellar parameters of the target star --- that is, after an appropriate burn-in phase. 
In Section~\ref{sec:hare}, we perform a hare and hound exercise based on another MCMC based code called Asteroseismic Inference on a Massive scale \citep[\textsc{aims},][]{Reese2016,LundReese2018,2019MNRAS.484..771R}. \textsc{aims} bypasses the high computational cost of MCMC by computing new samples by interpolation in an already existing grid of stellar models. Within a few hours, \textsc{aims} is thus able to investigate millions of a new combination of global stellar parameters and compare the stellar properties with observational constraints, mapping the posterior probability distribution. Like \textsc{hephaestus}, \textsc{aims} is based on the MCMC ensemble sampler by \cite{Goodman2010} using the implementation by \cite{emcee}.


\subsection{Method: Solar calibrations, likelihood, and priors} \label{sec:solar}

To produce a solar calibration model \textsc{garstec} uses a Newton solver to optimize for the structure model to fit observational constraints on the present-day Sun \citep{Weiss2008}. This is an iterative procedure: \textsc{garstec} computes several stellar evolution tracks of $1.0\,\mathrm{M}_\odot$ stars, adjusting the mixing length parameter ($\alpha_\textsc{mlt}$) and initial composition on the pre-main sequence (pre-MS), until the code recovers the solar luminosity ($L_\odot$), the solar radius ($R_\odot$), and the surface composition of the Sun at the present solar age. The result of this iterative calibration is a single structure model that recovers the required properties within a specified accuracy. While we thus arrive at a model of the present-day Sun, the Newton solver approach does not map the uncertainties of the global solar parameters into uncertainties on the properties of the {\color{black}final} structure model. To do so, we perform an MCMC analysis based on the same criteria as used in standard solar calibrations. In our analysis, we thus explore a three-dimensional parameter space, spanned by $\alpha_\textsc{mlt}$ as well as the initial hydrogen ($X_\mathrm{i}$), and heavy metal ($Z_\mathrm{i}$) abundances.

Like in a normal solar calibration, we keep the mass and the stellar age fixed to $1.0\,\mathrm{M}_\odot$ and $4.57\,$Gyr, respectively. Furthermore, we evaluate our model based on $L_\odot$, $R_\odot$, and the surface composition, i.e. $Z_\mathrm{S\odot}/X_\mathrm{S\odot}$. By only including these three observational constraints in our likelihood, we reliably map the uncertainties that are introduced when performing a standard solar calibration.
We thus vary three parameters ($X_\mathrm{i}$, $Z_\mathrm{i}$, and $\alpha_\textsc{mlt}$) to fit three observables ($L_\odot$, $R_\odot$, and $Z_\mathrm{S\odot}/X_\mathrm{S\odot}$).

To facilitate an easy comparison with the literature, we use the same constraints on $L_\odot$ as \cite{Bahcall2006}. As regards the solar radius, we draw upon \cite{Brown1998}. We use AGSS09 and set the uncertainty on $\mathrm{[Fe/H]}$ to be $0.05\,\mathrm{dex}$. This is equivalent to the uncertainties of the most abundant metals as well as on iron.

We employ broad uniform priors for all three parameters. Since a solar calibration based on coupled stellar models from \textsc{garstec} yields a mixing length parameter of 4.9 \citep{JoergensenWeiss2019}, we restrict ourselves to map the parameters space for $\alpha_\mathrm{mlt}$ between 4.0 and 8.0. As regards the initial chemical composition, we require that the initial helium content is larger or equal to the primordial value from big bang nucleosynthesis (i.e. $Y_i \geq Y_{BBN}=0.245$, \citealt{Planck2016}).
The discussed observational constraints are listed in the upper panel of Table~\ref{tab:paramsun}.

\begin{table}
    \centering
    \caption{Summary of our solar MCMC analysis. The uppermost three rows contain the employed constraints. The lowermost rows contain a summary of the posterior probability distributions of the obtained stellar parameters, including the median and the percentiles of the $68\%$ credibility interval. We set $\mathrm{Z_\mathrm{S\odot}/X_\mathrm{S\odot}}=0.0180$.}
    \label{tab:paramsun}
    \begin{tabular}{lccccccccccccccccccccc} 
        \hline
        $R$ & $695,508 \pm 26^a\,$km \\ [2pt]
        $\mathrm{[Fe/H]}$ & $0.00\pm0.05^b$ \\ [2pt]
        $L$ & $(3.842\pm 0.0154^c)\times 10^{33} \,\mathrm{erg\, s^{-1}}$ \\ [2pt]
        \hline
        $\alpha_\textsc{mlt}$ & $5.77^{+ 0.99}_{-0.71}$ \\ [2pt]
        $X_\mathrm{i}$ & $0.7203^{+ 0.0076}_{-0.0069}$ \\ [2pt]
        $Z_\mathrm{i}$ & $0.01497 ^{+ 0.00096}_{-0.00104}$ \\ [2pt]
        \hline
\end{tabular}
\flushleft \item[] $^a$ Seismic constraint by \cite{Brown1998}.
\item[] $^b$ Based on the composition by \cite{Asplund2009}.
\item[] $^c$ Constraint from \cite{Bahcall2006} and references therein.
\end{table}


\subsection{Results and discussion} \label{sec:solar2}

We have performed an analysis with 32 walkers, accumulating 7488 samples, after discarding a burn-in phase. The obtained posterior probability distributions on $\alpha_\textsc{mlt}$, $X_\mathrm{i}$, and $Z_\mathrm{i}$ are summarized alongside the observational constraints in Table~\ref{tab:paramsun}.

We note that our analysis yields a broad posterior probability distribution for $\alpha_\textsc{mlt}$. This is consistent with an analysis of Alpha Centauri A and B, for which \cite{JoergensenAngelou2019} found that the structure and evolution of our coupled models are rather insensitive to the value taken by $\alpha_\textsc{mlt}$. This is because the mixing length parameter only dictates the structure of a narrow nearly-adiabatic layer, as discussed in Section~\ref{sec:method}.

\begin{figure}
\centering
\includegraphics[width=1.0\linewidth]{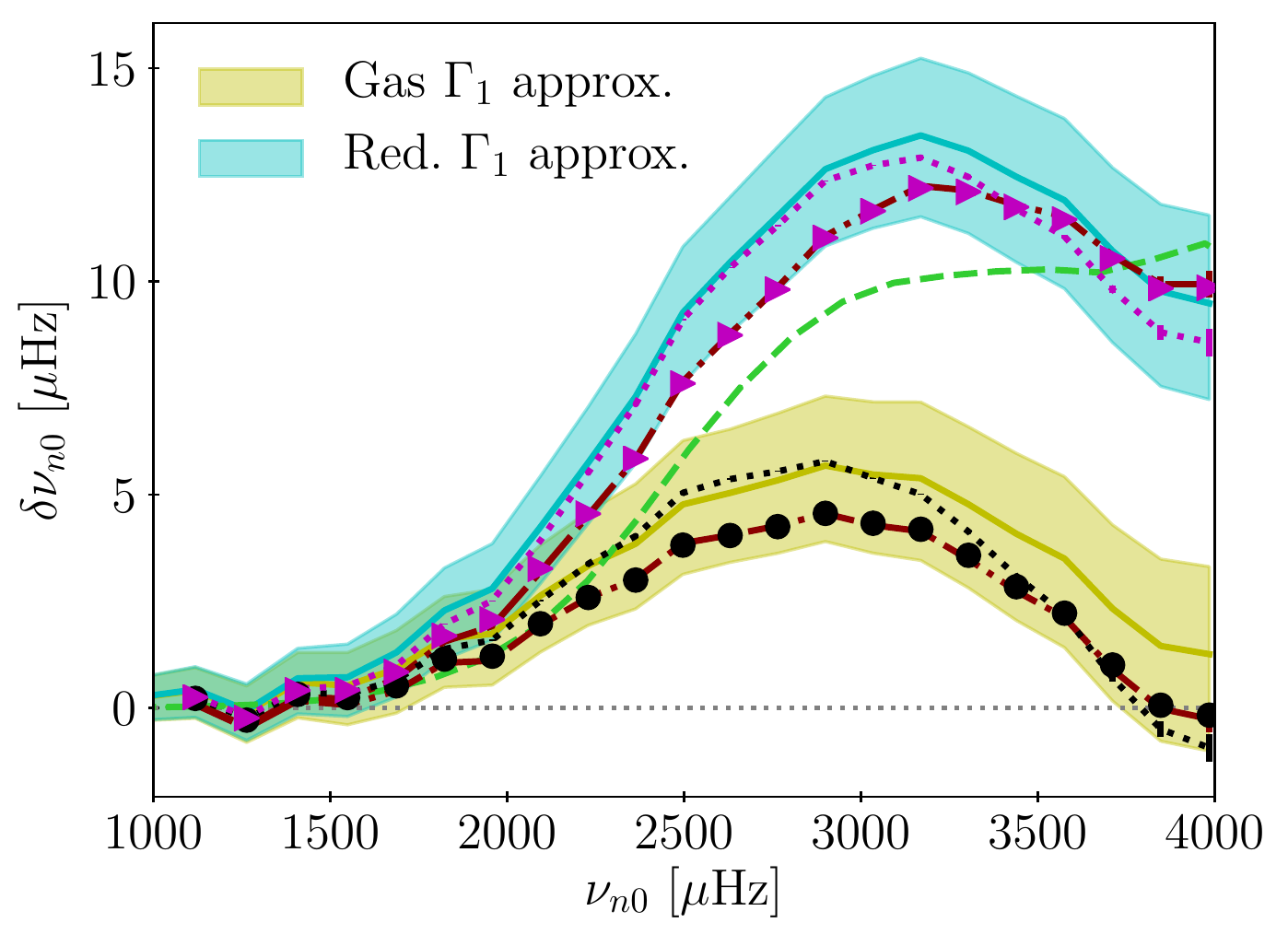}
\caption{$68\,\%$ credibility intervals on the residuals $\delta \nu_{n\ell}$) between the solar model frequencies and BiSON observations within the gas $\Gamma_1$ (yellow) and reduced $\Gamma_1$ (cyan) approximations. The corresponding solid line indicates the median. The plot is based on the 7488 samples from our MCMC analysis. The dashed green line shows the modal effect computed by \citet{Houdek2017} for the present-day Sun (courtesy of G.~Houdek). The dots and triangles show the results obtained from a solar calibration using the \textsc{cl{\'e}s} stellar evolution code (case a in Table~\ref{tab:solcaltab}). The dash-dotted dark red lines indicate the results that were obtained from the \textsc{garstec} solar calibration model presented by \citet{JoergensenWeiss2019}. Finally, the dotted black and purple line indicate the frequencies obtained when increasing the solar radius for the \textsc{cl{\'e}s} model by $6\,$km (case b in Table~\ref{tab:solcaltab}).}
\label{fig:Sunfreqdiff}
\end{figure}

\begin{figure}
\centering
\includegraphics[width=1.0\linewidth]{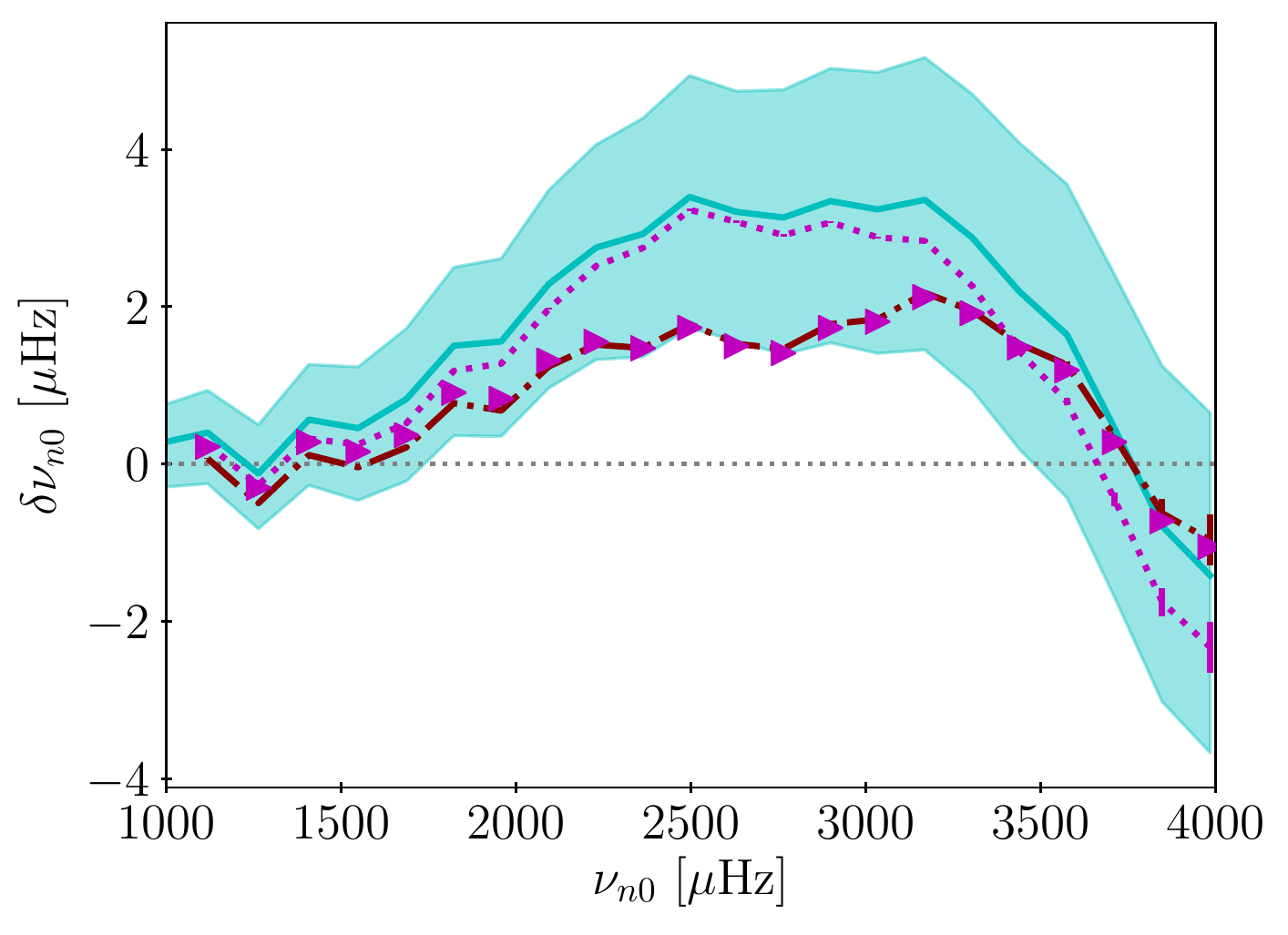}
\caption{Pendant to Fig.~\ref{fig:Sunfreqdiff} including only the frequencies computed within the reduced $\Gamma_1$ approximation after subtracting the modal contribution to the surface effect.
}
\label{fig:Sunfreqdiff2}
\end{figure}

We computed stellar oscillations for all 7488 realisations of the present-day Sun in our MCMC analysis. This allowed us to construct the posterior probabilities of the model frequencies. Figure~\ref{fig:Sunfreqdiff} shows a comparison between the resulting posterior distributions and observations from the Birmingham Solar Oscillation Network \citep[BiSON: ][]{Broomhall2009, Davies2014} within the gas and reduced $\Gamma_1$ approximations. Figure~\ref{fig:Sunfreqdiff} also includes the modal effect determined by \cite{Houdek2017}. To include non-adiabatic effects in the comparison between the adiabatic model frequencies and observations, one simply has to subtract these modal effects from the model frequencies within the reduced $\Gamma_1$ approximation {\color{black}\citep[cf.][]{Houdek2017}}. This is illustrated in Fig.~\ref{fig:Sunfreqdiff2}. We note that we do not include any uncertainties on the modal surface effect. This is because such error bars are currently not available and because the computation of such uncertainties lies beyond the scope of this paper.

In addition to the results of the MCMC analysis, Figs~\ref{fig:Sunfreqdiff} and \ref{fig:Sunfreqdiff2} include the results from the solar calibration by \cite{JoergensenWeiss2019} as well as two solar calibration models that have been computed using the \textsc{cl{\'e}s} stellar evolution code. We summarize key numbers for these solar calibrations in Table~\ref{tab:solcaltab}. 

\begin{table}
    \centering
    \caption{Summary of solar calibrations. All models are calibrated to recover the solar radius by \citet{Brown1998} (695,508 km). For the \textsc{garstec} model, $1\,\mathrm{M_\odot}=1.9891\times 10^{33}\,$g, $G=6.6738\times10^{-8}\,\mathrm{cm^3\,g^{−1}\,s^{−2}}$, and $\mathrm{Z_\mathrm{S\odot}/X_\mathrm{S\odot}}=0.0179$. For the \textsc{cl{\'e}s} models, $1\,\mathrm{M_\odot}=1.9884\times 10^{33}\,$g, $G=6.6743\times10^{-8}\,\mathrm{cm^3\,g^{−1}\,s^{−2}}$, and $\mathrm{Z_\mathrm{S\odot}/X_\mathrm{S\odot}}=0.0181$. While \textsc{garstec} aims to fit an luminosity of $3.816\times10^{33}\,\mathrm{erg \, s^{-1}}$, \textsc{cl{\'e}s} aims for a luminosity of $3.828\times10^{33}\,\mathrm{erg \, s^{-1}}$.}
    \label{tab:solcaltab}
    \begin{tabular}{lccccccccccccccccccccc} 
        \hline
        Model & $R$ [km] & $L$ [erg s$^{-1}$] & $\alpha_\mathrm{mlt}$ & $Z_\mathrm{i}$ & $X_\mathrm{i}$ \\ [2pt]
        \hline
        \textsc{garstec} & $695,494$ & $3.816\times 10^{33}$ & 4.876 & 0.0149 & 0.7215 \\ [2pt]
        \textsc{cl{\'e}s} (a) & $695,565$ & $3.830\times 10^{33}$ & 3.935 & 0.0151 & 0.7186 \\ [2pt]
        \textsc{cl{\'e}s} (b) & $695,571$ & $3.830\times 10^{33}$ & 3.935 & 0.0151 & 0.7186 \\ [2pt]
        \hline
\end{tabular}
\end{table}

The \textsc{garstec} solar calibration model by \cite{JoergensenWeiss2019} recovers observations within $2\,\mu$Hz at all frequencies. One of the \textsc{cl{\'e}s} models (case a in Table~\ref{tab:solcaltab}) performs equally well, while the median \textsc{garstec} model from the MCMC analysis and the other solar \textsc{cl{\'e}s} model (case b in Table~\ref{tab:solcaltab}) yield slightly larger residuals. For comparison, the residuals of standard stellar models exceed the residuals shown in Fig.~\ref{fig:Sunfreqdiff2} by one order of magnitude (e.g. Model S, \citealt{jcd1996}).

The remaining residuals of our coupled models are still orders of magnitude larger than the measurement uncertainties. They may hence point towards missing input physics. For instance, as discussed by \cite{Magic2016}, the neglect of magnetic fields in the 3D simulations plays a role for the seismic properties of patched models. The same holds true for {\color{black}the} employed solar composition, the EOS, the opacity tables, and the boundary condition for the p-modes in the pulsation code. However, considering the inferred uncertainties on the model frequencies, we might at least partly account for the remaining residuals based on the uncertainties on the solar global parameters ($L$, $R$ and $Z_\mathrm{S}/X_\mathrm{S}$) alone. We also note that this finding brings the model frequencies of various patched models in the literature into line: while frequencies of published solar patched models differ by a few microhertz, this might at least partly reflect differences in the global stellar parameters. For further discussions on this topic, we refer the reader to \cite{Joergensen2017}, \cite{Joergensen2019}, and \cite{Schou2020}.

Furthermore, based on the same notion, we can explain the discrepancies between the different models in Figs~\ref{fig:Sunfreqdiff} and \ref{fig:Sunfreqdiff2}. For instance, the difference between the median of the MCMC run and the solar model presented by \cite{JoergensenWeiss2019} can be explained in terms of the difference in the adopted luminosity. The solar calibration model presented by \cite{JoergensenWeiss2019} is thus constructed assuming the solar luminosity to be $3.816\times10^{33}\,\mathrm{erg \, s^{-1}}$, in order to recover the effective temperature of the solar envelope simulation in the Stagger grid ($5768.5\,$K), while the solar luminosity used in the MCMC analysis is $3.842\times10^{33}\,\mathrm{erg \, s^{-1}}$. Similarly, the differences between the frequencies of the solar calibration model presented by \cite{JoergensenWeiss2019} and the \textsc{cl{\'e}s} solar calibration models can be explained in terms of the differences in the adopted luminosity and photospheric radius (cf. Table~\ref{tab:solcaltab}). The differences between the median MCMC model and the discussed solar calibrations are thus all well within the error bars that were determined by the MCMC {\color{black}analysis}.

\begin{figure}
\centering
\includegraphics[width=1.0\linewidth]{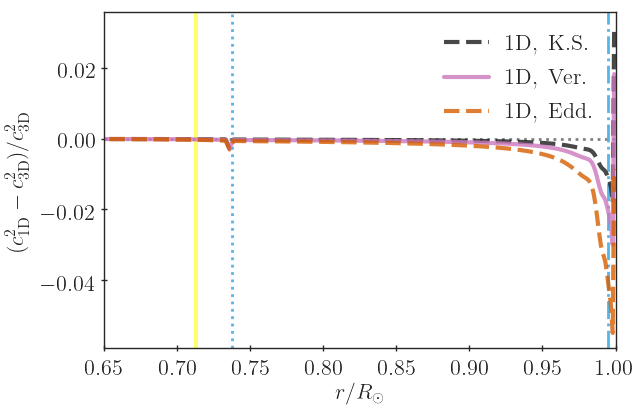}
\caption{{\color{black}Relative difference in the squared sound speed between a \textsc{cl{\'e}s} solar model and three standard solar models. The standard models employ an Eddington grey atmosphere (1D, Edd.) or the semi-empirical relations by \citet{KrishnaSwamy1966} (1D, K.S.) and \citet{1981ApJS...45..635V} (1D, Ver.). The deployed solar calibrations underlie the analysis in Section~\ref{sec:global}. In contrast to the solar models presented in Figs \ref{fig:Sunfreqdiff}, \ref{fig:Sunfreqdiff}, and \ref{fig:SunCdiff}, they do not include atomic diffusion. The dash-dotted cyan line at the right edge of the panel indicates the position of the lowermost meshpoint in the appended $\langle \mathrm{3D} \rangle$-envelope. The dotted cyan line shows the position of the lower convective boundary in the coupled model. The nearby small peak in the sound speed difference indicates a rather insignificant shift in the lower convective boundary that arises from the use of simple model atmospheres. The shaded yellow area indicates observational constraints on the lower convective boundary by \citet{Basu1997}. The discrepancy between model predictions and observations for the location of the lower convective boundary is a well-known tension that arises for AGSS09. For further comparisons between the structures of coupled and standard (solar) models, we refer the reader to Figs 2-4 in \citet{Joergensen2018}, Figs 2, 3, and 5 in \citet{JoergensenWeiss2019}, Figs 3 and 7 in \citet{JoergensenAngelou2019}, and Fig. 5 in \citet{Mosumgaard2020}. For a detailed depiction of the interior structure of coupled models, we refer the reader to Figs 1 and 10 in \citet{JoergensenAngelou2019}.}
}
\label{fig:comparison2}
\end{figure}

Finally, we turn to a discussion on the interior solar structure. The deep adiabat of the Sun, i.e. the entropy in solar adiabatic convective zone, is determined by the global solar parameters. It is, therefore, almost fully independent of whether we append a $\langle 3 \mathrm{D }\rangle$-envelope or use a standard 1D atmosphere to set the outer boundary conditions {\color{black}(cf. Fig.~\ref{fig:comparison2})}. This is not to say that the improved boundary conditions do not affect the structure below the appended $\langle 3 \mathrm{D }\rangle$-envelope. Indeed, as discussed by \cite{JoergensenWeiss2019}, the use of coupled models improves the overall sound speed profile {\color{black}in the upper convective layers (cf. Fig.~\ref{fig:comparison2})}. Meanwhile, the use of $\langle 3 \mathrm{D }\rangle$-envelopes {\color{black}as} the upper boundary conditions does, for instance, not affect the location of the base of the convective envelope {\color{black}significantly}. Indeed, the depth of the convection zone relative to the solar radius is rather insensitive to the adiabat for a fixed equation of state and fixed opacity tables, as discussed by \cite{JCD1997a}.

\begin{figure}
\centering
\includegraphics[width=1.0\linewidth]{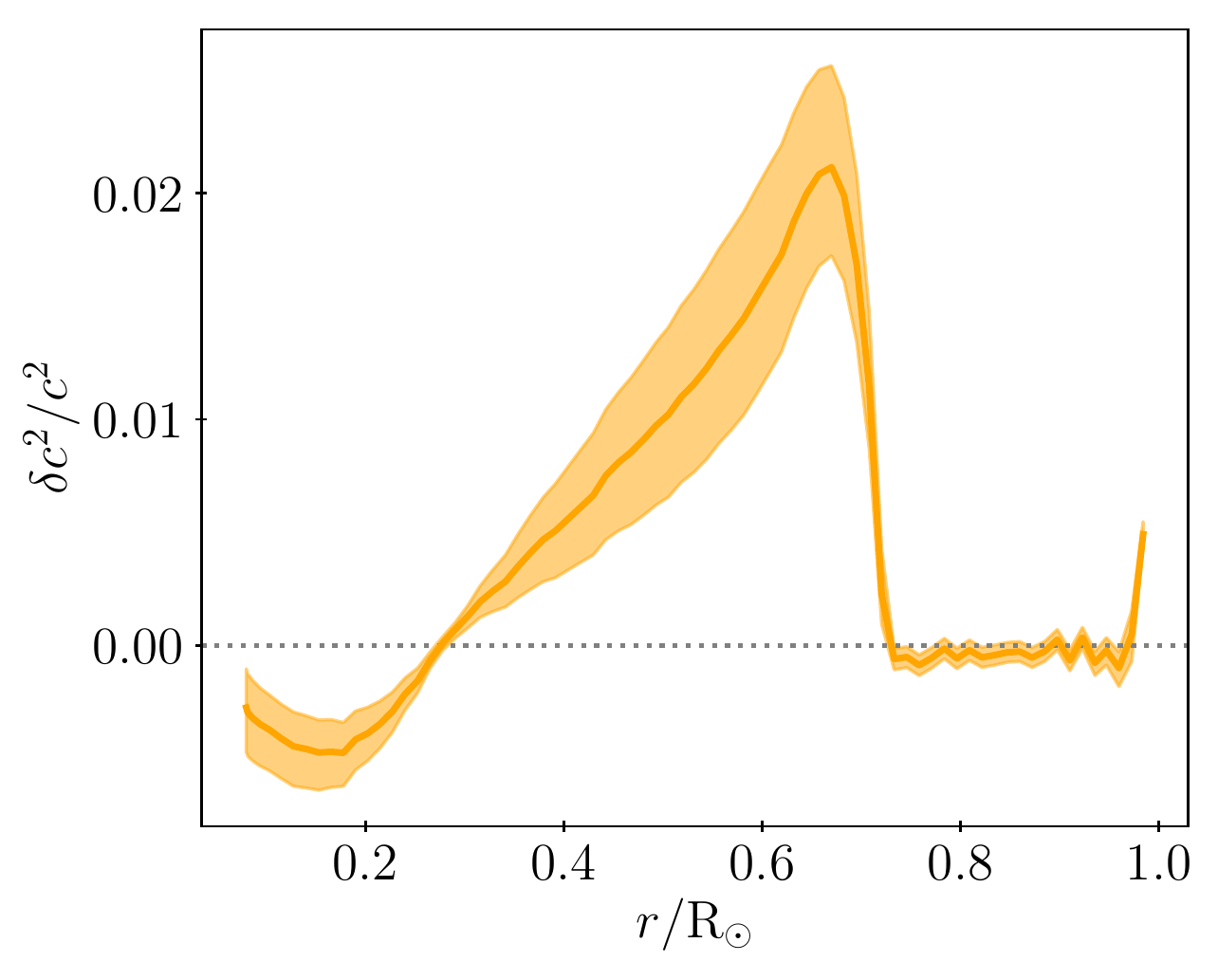}
\caption{Squared difference between the sound speed profile of the solar model and the inferred sound speed by \citet{Basu2008}. The shaded area shows the $68\,\%$ credibility intervals that we infer from our MCMC analysis. The solid line indicates the median of the sound speed profile distribution. The uncertainties in the plot solely refer to the uncertainties introduced by the global stellar parameters. For a discussion on the errors that stem from the inversion method, we refer to \citet{Degl1997} and \citet{Vinyoles2017}. 
}
\label{fig:SunCdiff}
\end{figure}

In this paper, we employ AGSS09. As shown by \citep{Serenelli2009}, this composition leads to a particularly strong disagreement with observations near the base of the convective envelope: the sound speed profiles of the stellar models are incompatible with the sound speed profile inferred from helioseismic constraints. The use of $\langle 3 \mathrm{D }\rangle$-envelopes does not solve this shortcoming. Indeed, while the use of $\langle 3 \mathrm{D }\rangle$-envelopes makes patched models and our coupled models superior to {\color{black}standard} stellar models, the improved outer boundary conditions do not solve all tensions with seismic measurements. We illustrate this for the sound speed profile in Fig.~\ref{fig:SunCdiff}. The tension at the lower boundary of the convection zone may, however, be addressed by including overshooting \citep[e.g.][]{Schlattl1999,Baraffe2017,JoergensenWeiss2018} or altering the opacities {\citep[e.g.][and references therein]{JCD1997b,JCD1997a,Montalban2004,Montalban2006,2009A&A...494..205C,2018MNRAS.477.3845C}}. Finally, earlier measurements of the solar mixture \citep{Grevesse1993} lead to better agreement with helioseismology. For a {\color{black}recent} discussion of this pending issue, we refer the reader to \cite{2019FrASS...6...42B}.

\section{The gas $\Gamma_1$ approximation} \label{sec:gasG1}

As can be seen from the solar models presented above, the gas $\Gamma_1$ approximation recovers observations within a few microhertz in the case of the present-day Sun. While this still corresponds to several standard deviations of the observed frequencies, it is a sizeable improvement over the uncorrected frequencies of standard stellar models. Many authors have, therefore, assumed that the gas $\Gamma_1$ approximation performs reasonably well across the HR diagram. However, there is no justification for this approach beyond the fact that it yields reasonable results for the present-day Sun.

{\color{black}As shown by \cite{Houdek2017}, the reduced $\Gamma_1$ approximation appropriately accounts for the \textit{adiabatic} contribution of the turbulent pressure to the eigenfrequencies in the case of the present-day Sun. We} can thus recover the observed frequencies by computing the adiabatic frequencies within the \textit{reduced} $\Gamma_1$ approximation and subsequently adding the modal {\color{black}effect} (cf. Fig.~\ref{fig:Sunfreqdiff2}). It follows that the difference between the reduced and gas $\Gamma_1$ approximations should correspond to the modal effect across the HR diagram, if the gas $\Gamma_1$ approximation indeed recovers the observed frequencies for stars other than the Sun, {\color{black}and if the assumptions that underlie the reduced $\Gamma_1$ approximation hold true for these stars}. 
To establish the validity of the gas $\Gamma_1$ approximation, we, therefore computed the frequency difference between the gas and reduced $\Gamma_1$ approximations at the frequency of maximum power ($\nu_\mathrm{max}$) for different stellar parameters. We then compared this difference with the modal effect at $\nu_\mathrm{max}$ presented in Fig. 5 of \cite{Houdek2019}. {\color{black}Note that the modal effect presented by \cite{Houdek2019} has been computed from fully non-adiabatic calculations by subtracting adiabatic frequencies that were computed within the reduced $\Gamma_1$ approximation. By comparing the difference between the reduced and gas $\Gamma_1$ approximations to the results in \cite{Houdek2019}, we are thus directly comparing the gas $\Gamma_1$ approximation to the outcome of a fully non-adiabatic time-dependent treatment. The inferred \textit{absolute} errors of the gas $\Gamma_1$ approximation do hence not depend on the validity of the assumptions that underlie the reduced $\Gamma_1$ approximation.}

For the computation of $\nu_\mathrm{max}$, we adopt \citep{Brown1991}:
\begin{equation}
\nu_\mathrm{max} = \left(\frac{M}{\mathrm{M}_\mathrm{\odot}}\right)\left(\frac{R}{\mathrm{R}_\mathrm{\odot}}\right)^{-2}\left(\frac{T_\mathrm{eff}}{\mathrm{T}_\mathrm{eff\odot}}\right)^{-1/2}\nu_\mathrm{max,\odot} , \label{eq:numax1}  
\end{equation}
where $\nu_\mathrm{max\odot} = 3090\,\mu$Hz \citep{2011ApJ...743..143H, Hekker2020}, and $T_\mathrm{eff\odot}=5777\,$K.

We have computed the model frequencies within the gas and reduced $\Gamma_1$ approximation for a grid of coupled models of main-sequence stars with effective temperatures between 5750 and 6700$\,$K and with $\log g$ between 4.0 and 4.5 dex. We hereby cover the same region of the Kiel diagram as explored by \cite{Houdek2019}. All models in the grid are computed without diffusion so that models that enter the analysis have solar metallicity ($\mathrm{[Fe/H]}=0.0$). The composition is based on the solar abundances evaluated by \cite{Asplund2009} and a solar calibration that was likewise performed without including diffusion ($X_\mathrm{i}=0.7301$, $Z_\mathrm{i}=0.013214$, and $\alpha_\textsc{mlt}=1.82$).

\begin{figure}
\centering
\includegraphics[width=1.0\linewidth]{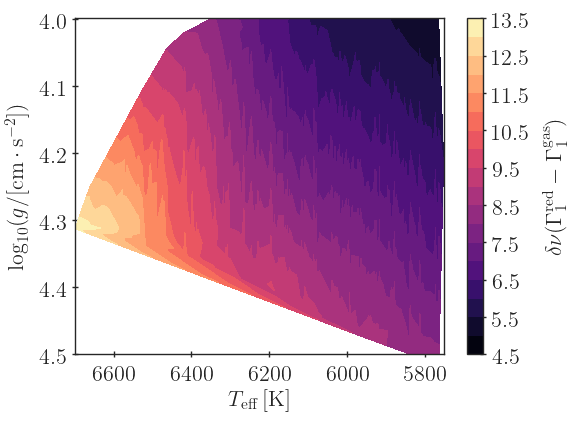}
\includegraphics[width=1.0\linewidth]{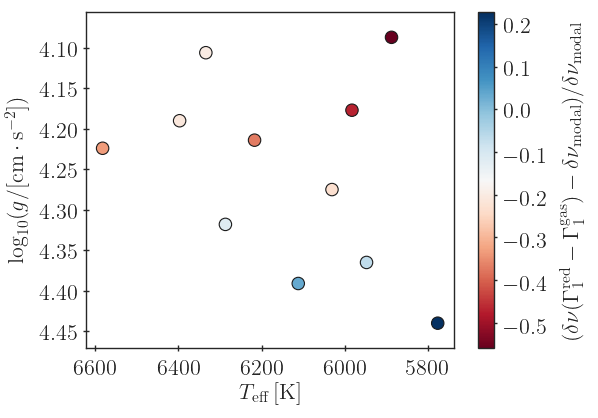}
\includegraphics[width=1.0\linewidth]{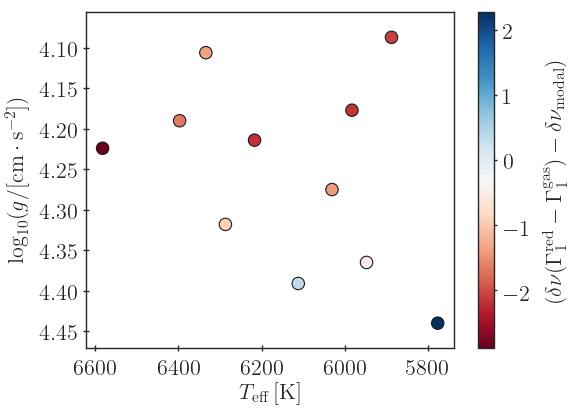}
\caption{\textbf{Upper panel:} Kiel diagram, showing the frequency difference between the reduced and the gas $\Gamma_1$ approximations at $\nu_\mathrm{max}$. The plot is based on 418 stellar structure models. \textbf{Middle panel:} Here, we have subtracted the modal effect determined by \citet{Houdek2019} from the results shown in the upper panel and plotted the relative differences. {\color{black}\textbf{Lower panel:} Pendant to middle panel but showing absolute rather than relative errors.}}
\label{fig:GGvsRG}
\end{figure}

The frequency difference between the reduced and gas $\Gamma_1$ approximations at $\nu_\mathrm{max}$ are shown in the upper panel of Fig.~\ref{fig:GGvsRG}. From a qualitative comparison with the results in the paper by \cite{Houdek2019}, one can see that the difference between the reduced and the gas $\Gamma_1$ approximations show the same overall trends across the Kiel diagram as the modal effect does. Combining our results with those listed in Tables 1 and 2 in the paper by \cite{Houdek2019}, we find that the difference between the reduced and gas $\Gamma_1$ approximations recovers the modal surface effect within $50\,\%$ of the modal effect across the sampled region of the parameter space. {\color{black}The corresponding absolute error that results from the use of the gas $\Gamma_1$ approximation is thus at most $2.9\,\mu\mathrm{Hz}$ across the explored region of the parameter space.}
These findings are illustrated in the two lower panels of Fig.~\ref{fig:GGvsRG}.

While a discrepancy of {\color{black}up to $2.9\,\mu$Hz (or} $50\,\%$) is substantial, we note that the gas $\Gamma_1$ approximation recovers the solar observations with a similar accuracy (cf. Fig.~\ref{fig:Sunfreqdiff}). Indeed, for a large fraction of the sampled parameter space, the gas $\Gamma_1$ approximation even performs better at $\nu_\mathrm{max}$ than in the case of the Sun. We thus conclude that the gas $\Gamma_1$ approximation performs as well for other low-mass main-sequence stars as it does for the Sun. {\color{black}While 
we are thus able to demonstrate the fitness of the gas $\Gamma_1$ approximation beyond the Sun, we do not directly provide a physical justification for the underlying assumptions. Rather, our results indirectly gain a physical justification through the fully non-adiabatic calculations by \cite{Houdek2019}, to which we compare.}


\section{Global stellar properties} \label{sec:global}

\begin{figure}
\centering
\includegraphics[width=1.0\linewidth]{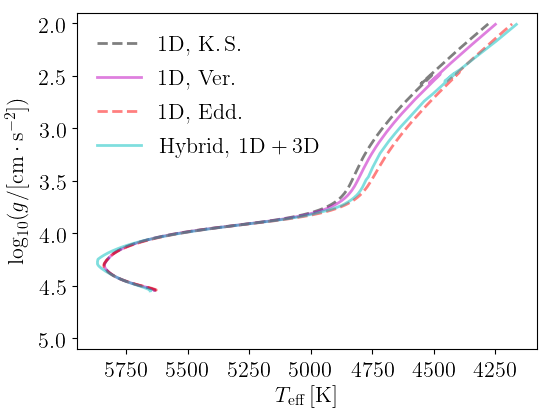}
\caption{Kiel diagram, showing the theoretical stellar evolution track of a $1\,\mathrm{M_\odot}$ star using different outer boundary conditions. Here, we include our coupled stellar models (Hybrid, 1D+3D) as well as standard stellar models that employ Eddington grey atmospheres (1D, Edd.) or the semi-empirical relations by \citet{KrishnaSwamy1966} (1D, K.S.) and \citet{1981ApJS...45..635V} (1D, Ver.).}
\label{fig:Sun_trackdiff}
\end{figure}

As discussed by \cite{JoergensenWeiss2019}, \cite{JoergensenAngelou2019}, and \cite{Mosumgaard2020}, the use of $\langle \mathrm{3D} \rangle$-envelopes as the outer boundary layers affects the predicted stellar evolution tracks. This is illustrated in Fig.~\ref{fig:Sun_trackdiff} for a $1\,\mathrm{M_\odot}$ star. The figure includes the evolution of a coupled stellar model {\color{black}as well as of three standard stellar models. The standard models are based on} different $T(\tau)$-relations that are commonly found in the literature: Eddington grey atmospheres and the semi-empirical relations by \cite{KrishnaSwamy1966} and \cite{1981ApJS...45..635V}. {\color{black}Each} of the stellar evolution tracks in Fig.~\ref{fig:Sun_trackdiff} passes through the present-day Sun by default. Each track is thus based on a distinct solar calibration, for which we employ the same outer boundary conditions. As can be seen from the figure, the use of $\langle \mathrm{3D} \rangle$-envelopes affects both the predicted turn-off point {\color{black}(TO)} and the evolution on the RGB.

\begin{figure}
\centering
\includegraphics[width=1.0\linewidth]{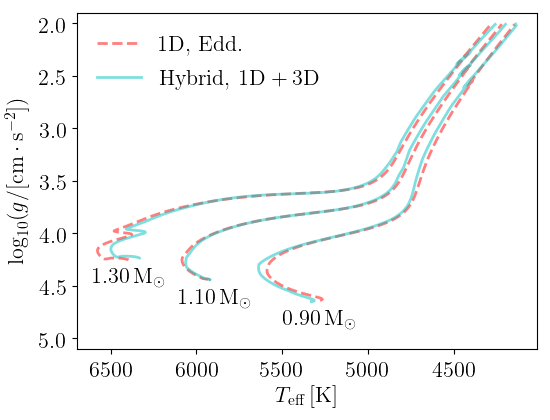}
\includegraphics[width=1.0\linewidth]{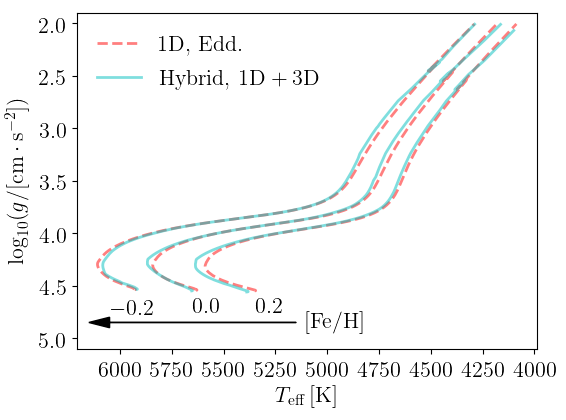}
\caption{\textbf{Upper panel:} Kiel diagram, showing the theoretical stellar evolution track of a $0.90\,\mathrm{M_\odot}$, a $1.10\,\mathrm{M_\odot}$, and a $1.30\,\mathrm{M_\odot}$ star using different outer boundary conditions. Here, we include our coupled stellar models (Hybrid, 1D+3D) as well as standard stellar models that employ Eddington grey atmospheres (1D, Edd.). {\color{black}\textbf{Lower panel:} Evolution tracks of $1.00\,\mathrm{M_\odot}$ standard and coupled models for different metallicities.}}
\label{fig:tracks2}
\end{figure}

In this section, we further quantify {\color{black}the impact of $\langle \mathrm{3D} \rangle$-envelopes on the inferred global stellar properties by comparing our coupled models with standard stellar models at different masses, ages and metallicities. Based on} Fig.~\ref{fig:Sun_trackdiff}, we note that the simple Eddington grey atmosphere does a better job than its semi-empirical counterparts at recovering the evolution of the coupled models. We, therefore, perform the majority of the following comparisons, using standard stellar models that employ Eddington grey atmospheres. A selection of the resulting evolution tracks are shown in Fig.~\ref{fig:tracks2}.

Other authors have included information from 3D simulations into stellar evolution codes by varying $\alpha_\mathrm{mlt}$ across the Kiel diagram \citep[][{see also Appendix~\ref{sec:tayar}}]{Trampedach2014b, Magic2015}. It is worth noting that the resulting changes in the stellar evolution tracks are qualitatively consistent with the results presented in Fig.~\ref{fig:Sun_trackdiff} \citep[see][for a more detailed discussion]{Mosumgaard2020}. Both \cite{Mosumgaard2018} and \cite{Sonoi2019} thus find that the predicted variation in $\alpha_\mathrm{mlt}$  leads to higher effective temperatures on the RGB {\color{black}than} standard stellar models with constant $\alpha_\mathrm{mlt}$ and Eddington grey atmospheres (cf. Figs 3 and 4 in \cite{Mosumgaard2018} and Fig. 15 in \cite{Sonoi2019}).

Meanwhile, we note that the use of coupled models leads to a shift in the {\color{black}TO} that is not observed when using a variable mixing length parameter \citep[e.g.][]{Mosumgaard2018,Sonoi2019}. This might imply that the resolution of the Stagger grid is too low in the corresponding region of the HR diagram for {\color{black}our interpolation scheme to perform well (cf. Appendix~\ref{sec:interr})}. If so, the position of the {\color{black}TO} for our coupled models might be subject to interpolation errors. On the other hand, we note that the use of a varying mixing length parameter comes with its own caveats. Firstly, the varying mixing length parameter is calibrated based on the existing 3D RHD simulations and is then varied across the HR diagram by interpolation in these calibrated values. The varying mixing length parameter approach is itself thus subject to the assumptions that enter through the chosen interpolation algorithm and the low resolution of the underlying grids. Secondly, it has been shown by e.g. \cite{Trampedach2011} that the mixing length parameter not only varies as a function of the global stellar parameters but also as a function of depth. The use of a constant mixing length parameter throughout the interior structure is thus a simplifying assumption. The procedures by \cite{Mosumgaard2018} and \cite{Sonoi2019}
do not account for this {\color{black}and do hence not recover the stratification of the underlying 3D simulations \citep[cf.][]{Joergensen2017,Mosumgaard2018}. Meanwhile}, as shown by \cite{Sonoi2019}, a shift near the {\color{black}TO} similar to that in Fig.~\ref{fig:Sun_trackdiff} appears between tracks computed using MLT and FST \citep[see also][] {Mazzitelli1995,2002ApJ...564L..93D}. Since the stratification predicted by MLT and FST are somewhat different, the finding by \cite{Sonoi2019} tells us that a shift in the {\color{black}TO} may arise, if the variation of the mixing length parameter with depth changes throughout the HR diagram --- in this picture, from a more MLT-like to a more FST-like behaviour. In this scenario, the shift in the {\color{black}TO} that arises from the use of coupled models (cf. Fig.~\ref{fig:Sun_trackdiff}) might be a physical feature rather than stemming from an interpolation error. To shed light on this issue, however, further 3D simulations are needed. This is beyond the scope of this paper. We thus restrict ourselves to raise caution regarding the behaviour of our coupled models near the {\color{black}TO. However, we also note that any interpolation errors that might occur at the TO neither affect the previous nor the subsequent evolution of the stellar models.}

All models in this Section were computed using the \textsc{cl{\'e}s} stellar evolution code. They have all been computed without atomic diffusion, in order to ensure a constant metallicity along the stellar evolution tracks. Our coupled models include turbulent pressure in the appended $\langle \mathrm{3D} \rangle$-envelope, while we ignore the contribution for turbulent pressure to hydrostatic equilibrium the deep interior. In contrast, the \textsc{garstec} models in Section~\ref{sec:solar} include turbulent pressure throughout the stellar structure calibrated based on the appended $\langle \mathrm{3D} \rangle$-envelopes \citep{JoergensenWeiss2019}. This being said, the contribution of the turbulent pressure to the total pressure is small below the $\langle \mathrm{3D} \rangle$-envelope compared to its contribution within the envelope. Furthermore, as shown by \cite{JoergensenWeiss2019}, the stellar evolution track would be left unaffected, even if the turbulent pressure were to be ignored altogether (cf. Fig.~7 in \citealt{JoergensenWeiss2019}).


\subsection{Comparing models at solar metallicity} \label{sec:FeH0} \label{sec:globalFeH0}

{\color{black}In this section, we investigate  stars at solar metallicity}. For this purpose, we constructed a grid of coupled models with masses between 0.88 and $1.32\,\mathrm{M}_\odot$ with a step-size of $0.01\,\mathrm{M}_\odot$. For all models, $\mathrm{[Fe/H]}=0$. The resulting stellar evolution tracks are illustrated in Fig.~\ref{fig:tracks} in Appendix~\ref{sec:interr}. A subsample of structure models in this grid is used in the analysis presented in Section~\ref{sec:gasG1}. 
For comparison, we have constructed a grid of standard stellar models with masses between 0.80 and $1.50\,\mathrm{M}_\odot$ with a step-size of $0.01\,\mathrm{M}_\odot$. Again, we only include models, for which $\mathrm{[Fe/H]}=0$.

\begin{figure}
\centering
\includegraphics[width=1.0\linewidth]{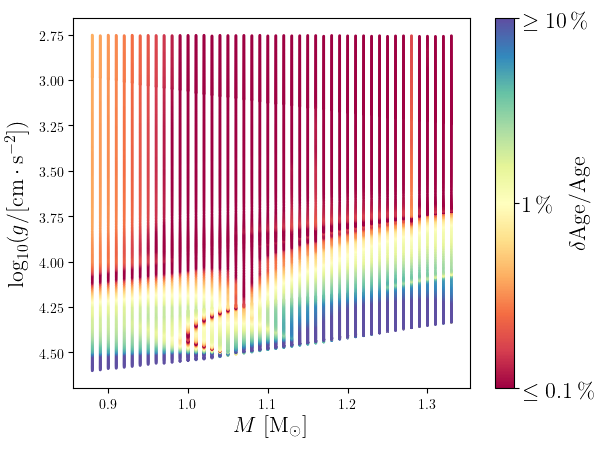}
\caption{Relative difference in age between coupled and standard stellar models with the same mass and radius. All models were computed at solar metallicity. For clarity, we have limited the colour scale to distinguish models, for which the age difference lies at or above $10\,\%$ and at or below $0.1\,\%$.
}
\label{fig:Agediff}
\end{figure}

For main-sequence stars, we find that the predicted ages are strongly affected by the outer boundary conditions when considering fixed masses and radii. 
{\color{black}Across the main sequence, the age differences lie close to or even exceed the 10 per cent accuracy.} If one were to infer the stellar age based on tight constraints on the stellar mass and radius, coupled stellar models would hence lead to different age estimates than their standard stellar counterparts. This finding is illustrated in Fig.~\ref{fig:Agediff}.

\begin{figure}
\centering
\includegraphics[width=1.0\linewidth]{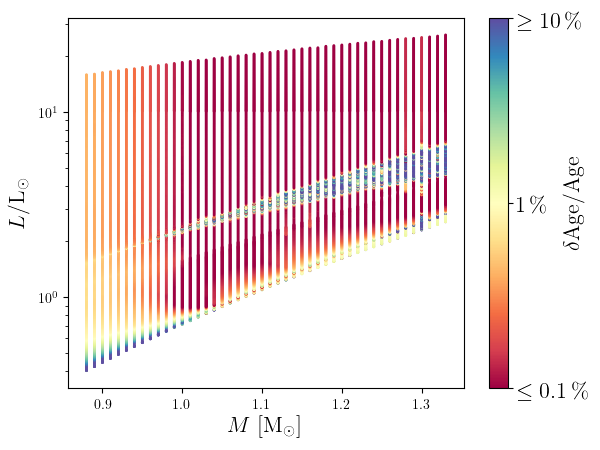}
\caption{Pendant to Fig.~\ref{fig:Agediff}: Relative difference in age between coupled and standard stellar models with the same mass and luminosity.
}
\label{fig:Agediff_lum}
\end{figure}

{\color{black}If we instead compare coupled and standard models with the same mass and luminosity, we again find that the largest discrepancies in age occur on the main sequence and near the {\color{black}TO}. This is illustrated in Fig.~\ref{fig:Agediff_lum}. However, in this comparison, the age discrepancy is rather low for a large fraction of main-sequence stars. In accordance with this, \cite{JoergensenAngelou2019} and \cite{Mosumgaard2020} both find that asteroseismic analyses based on both coupled and standard stellar models indeed yield mutually consistent age estimates for target stars on the main sequence when choosing a suitable likelihood. We thus} note that the established age difference arises from changes in the properties, based on which the age is pinned down. The outer boundary conditions do not fundamentally change the evolutionary timescales. For a star with a given mass and chemical composition the age is hence largely independent on the boundary conditions.

{\color{black}Independently of the parameters that enter our comparison, the same age is obtained for standard and coupled \textit{solar} models since both grids are based on solar calibrations.} The two underlying solar calibrations yield the same initial hydrogen and heavy metal abundance within $\times 10^{-4}$ and $\times 10^{-6}$, respectively. The calibrated mixing length parameter is 1.67 and 1.82 for the standard and coupled models, respectively. We use the values from the solar calibrations throughout the respective grids but note that this is a simplifying approximation (cf. Appendix~\ref{sec:tayar}). Nevertheless, this assumption is commonly used, and adopting it thus allows for a point of comparison with the literature.

On the RGB, we find that the absolute and relative discrepancy in age is much smaller than on the main sequence when comparing coupled and standard stellar with the same mass and radius {\color{black}(or luminosity)}. However, {\color{black}as} can be seen from Figs~\ref{fig:Sun_trackdiff} and \ref{fig:tracks2}, the effective temperature on the RGB as a function of the surface gravity is significantly altered by the {\color{black}use of} coupled stellar models. {\color{black}We thus find the age estimates of coupled and standard stellar models of RGB stars to differ substantially when comparing fixed positions in the Kiel diagram.}

\begin{figure}
\centering
\includegraphics[width=1.0\linewidth]{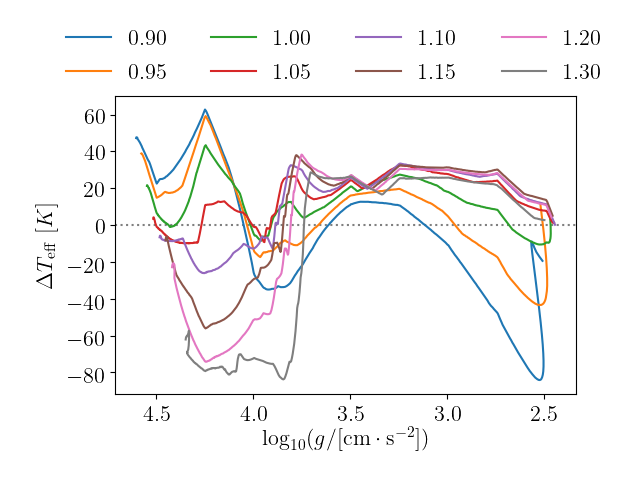}
\caption{Difference in effective temperature between standard and coupled stellar models (see also Fig.~\ref{fig:tracks2}). Positive residuals imply that the effective temperatures of the coupled models are higher than those of their standard model counterpart. We compare models with the same masses and ages. The mass is specified in the legend in units of the solar mass. On the abscissa, we specify the surface gravity of the coupled models, although we compare models at \textit{the same age}. The surface gravity of the corresponding standard model at the same age thus deviates from this value. We do so, in order to indicate the evolutionary phase of the star. For each star, we have included the evolution up until $\log g=2.5$.}
\label{fig:AgeTeff13}
\end{figure}

In Fig.~\ref{fig:AgeTeff13}, we compare standard and coupled models with the same mass and age {\color{black}to quantify the resulting} difference in the effective temperature. {\color{black}While} coupled models of RGB stars with low masses are found to be colder than their standard stellar counterparts, coupled models of RGB stars with masses above $1\,\mathrm{M_\odot}$ are warmer than the standard stellar models. The opposite is true on the main sequence.

\begin{figure}
\centering
\includegraphics[width=0.95\linewidth]{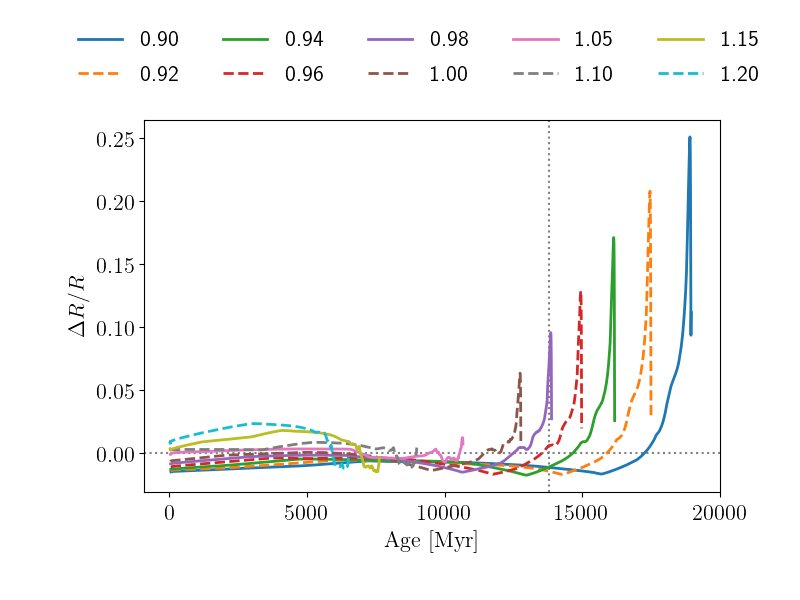}
\caption{Difference between the radius of standard stellar models and our coupled models as a function of age relative to the radius of the coupled model. Positive residuals imply that the coupled models have larger radii than their standard stellar counterparts. We compare models with the same masses and ages. The mass is specified in the legend in units of the solar mass. For each star, we have included the evolution up until $\log g=2.5$. The vertical dotted grey line indicates the present age of the Universe. The peak deviation is achieved close to $\log g=2.5$, corresponding to the red-giant luminosity bump \citep[cf.][]{2015MNRAS.453..666C}. At the same $\log g$, a decrease in $\Delta T_\mathrm{eff}$ occurs in Fig.~\ref{fig:AgeTeff13}.}
\label{fig:Rdiff}
\end{figure}

In Fig.~\ref{fig:Rdiff}, we likewise compare standard and coupled models with the same mass and age. Here, we investigate how the outer boundary conditions affect the predicted stellar radii. For models with masses below roughly $1\,\mathrm{M}_\odot$, we find large differences in the predicted stellar radii on the RGB. {\color{black}For} instance, the discrepancy in the predicted radius reaches $6\,\%$ for a $1.0\,\mathrm{M}_\odot$ star (at $\log g \approx 2.5$). This finding implies that standard stellar models of red giants attribute different seismic properties to the star than the corresponding coupled model with same age and mass would --- especially, for low masses. {\color{black}We discuss this further below.}

Based on Fig.~\ref{fig:Rdiff}, we note that the deviations in the stellar radius between coupled and standard stellar models are more complex than what one might anticipate based on patched stellar models. As discussed in the introduction, patched models are standard stellar models, for which the outermost layers are substituted by averaged RHD simulations \textit{after} computing the stellar evolution \citep[e.g.][]{Rosenthal1999}. Due to turbulent pressure and convective back-warming \citep{Trampedach2013, Trampedach2017}, 3D simulations of convective envelopes are more extended than their 1D counterparts. The radius of patched models thus always exceed that of the underlying standard stellar model. However, the improved boundary conditions do not leave the interior unaffected and alter the stellar evolution tracks. This is how a coupled stellar model can end up being smaller than a standard stellar model with the same mass and age.

\begin{figure}
\centering
\includegraphics[width=1.0\linewidth]{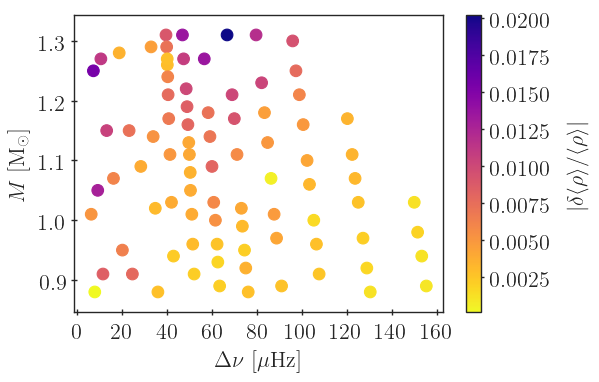}
\caption{Relative difference in the mean density between standard and coupled stellar models as a function of the large frequency separation and the mass for a selection of models at solar metallicity. The employed models have been chosen such that they cover the Kiel diagram in a regular pattern, i.e. such that different masses and all evolutionary stages are represented. We include coupled models for which $\log g \geq 2.75$. To ensure that we are comparing models with the exact same masses and large frequency separations, we interpolate in the corresponding global stellar parameters of the computed standard stellar models.
}
\label{fig:dnutest1}
\end{figure}

{\color{black}To illustrate how the use of coupled models affects the global seismic properties}, we computed the mean large separation ($\Delta \nu$) from the individual frequencies for a subset of coupled and standard stellar models:
\begin{equation}
\Delta \nu = \langle \nu_{n,\ell=0} - \nu_{n-1,\ell=0} \rangle.    
\end{equation}
We took the average over all frequencies between half and three halves of the frequency of maximum power ($\nu_\mathrm{max}$). {\color{black}In Fig.~\ref{fig:dnutest1}, we compare standard and coupled stellar models with the same mass and $\Delta \nu$. Because the evolution of coupled and standard stellar models differ, it stands to reason that standard and coupled models with the same mass and $\Delta \nu$ will differ in some other global properties. However, even without any impact of the outer boundary conditions on the predicted evolution tracks, the models would necessarily differ in other global properties. This is because the use of $\langle \mathrm{3D} \rangle$-envelopes partly mends the surface effect, which shifts the individual model frequencies and thus $\Delta \nu$. The model frequencies of the standard stellar models, on the other hand, have not been corrected to take the surface effect into account.
Indeed, we find the coupled models to have higher mean densities, as shown in Fig.~\ref{fig:dnutest1}.}

\begin{figure}
\centering
\includegraphics[width=1.0\linewidth]{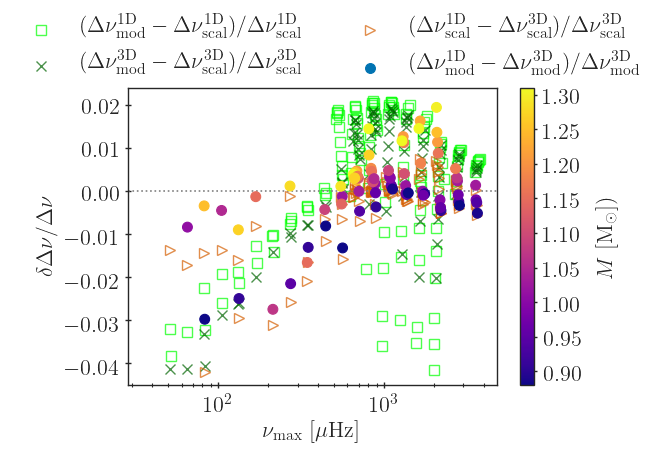}
\caption{Relative difference in the large frequency separation as a function of the frequency of maximum power. We include large frequency separations that are computed using the scaling relation given in Eq.~(\ref{eq:dnurel}) as well as from the individual frequencies of the models. These are denoted by the subscripts 'scal' and 'mod', respectively. At low (high) $\nu_\mathrm{max}$ the highest residuals stem from the models with lowest (highest) masses, i.e. those models, for which the highest residuals in the stellar radius is obtained (cf. Fig.~\ref{fig:Rdiff}).
}
\label{fig:dnutest2}
\end{figure}

{\color{black}Moreover, we find that the use coupled models leads to a higher value of $\Delta \nu$ when comparing coupled and standard stellar models with the same value of $\nu_\mathrm{max}$. Again, the explanation for this finding is twofold. First, the use of $\langle \mathrm{3D} \rangle$-envelopes partly mends the surface effect, shifting $\Delta \nu$. Secondly, while the coupled and standard stellar models that enter the comparison share the same $\nu_\mathrm{max}$, they do not share many other global properties. After all, $\nu_\mathrm{max}$ is computed based on Eq.~(\ref{eq:numax1}) and is thus sensitive to any shifts in mass, radius, and effective temperature between coupled and standard stellar models. If it is indeed the case that we are comparing models with different masses and radii, it follows that the standard and coupled stellar models in our comparison would also not lead to the same $\Delta \nu$ if we were to compute $\Delta \nu$ from a simple scaling relation \citep[e.g.][for a discussion on scaling relations]{Handberg2017, Rodrigues2017, Sahlholdt2018}}:
\begin{equation}
\Delta \nu_\mathrm{scal} = \Delta \nu_\odot \left(\frac{M}{\mathrm{M}_\odot}\right)^{1/2}
\left(\frac{R}{\mathrm{R}_\odot}\right)^{3/2}, \label{eq:dnurel}
\end{equation}
where we set $\Delta \nu_\odot = 135.1\,\mu$Hz \citep{2011ApJ...743..143H}. Indeed, when computing $\Delta \nu$ based on Eq.~(\ref{eq:dnurel}), we arrive at discrepancies in $\Delta \nu$ between the standard and coupled stellar models that are as large as the deviations in $\Delta \nu$ obtained from the individual frequencies. In both cases, the deviations between the standard and coupled models are thus of the order of $10^{-3}$ to $10^{-2}$ times $\Delta \nu$. {\color{black}We} show this in Fig.~\ref{fig:dnutest2}.

\subsection{Comparing models across metallicities} \label{sec:acrossFeH}

\begin{figure}
\centering
\includegraphics[width=1.0\linewidth]{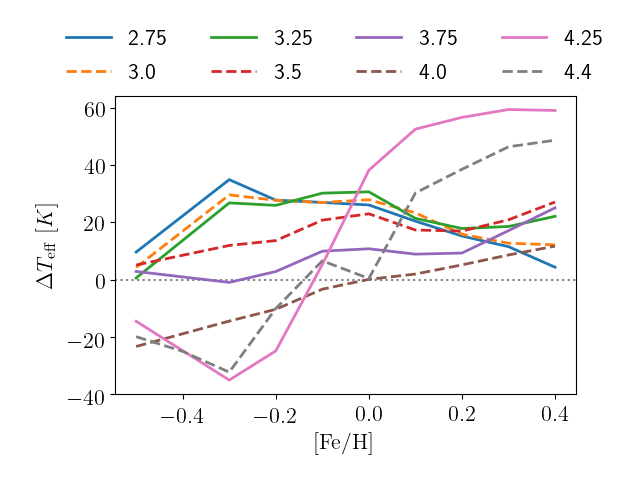}
\caption{Difference in effective temperature at fixed metallicities and surface gravities for stars with $1\,\mathrm{M}_\odot$. Positive residuals imply that the effective temperature of the coupled model is higher than that of its standard model counterpart. The surface gravity ($\log g$) is specified in the legend. To avoid complications arising from the red giant branch bump, we only include models for which $\log g >= 2.75$.}
\label{fig:Metals1}
\end{figure}

To evaluate the impact of metallicity on the conclusions drawn above, we have computed a set of coupled and standard stellar models with $\mathrm{[Fe/H]}$ between -0.5 and 0.4 in steps of 0.1. In all cases, we have fixed the stellar mass to $1.0\,\mathrm{M}_\odot$ and do not include diffusion. 
In Fig.~\ref{fig:Metals1}, we compare the effective temperature of standard and coupled models at different evolutionary stages. For this purpose, we compare structure models with the same surface gravity. At all metallicities, we find that our coupled stellar models yield higher effective temperatures on the RGB than the standard stellar models do ($\log g \leq 3.75$). The same conclusion is drawn for all masses at solar metallicity from Fig.~\ref{fig:Agediff} in Section~\ref{sec:FeH0}.

For the main-sequence, a more nuanced picture emerges. At super-solar metallicities, the coupled stellar models yield higher effective temperatures on the main-sequence and close to the {\color{black}TO} than the standard stellar models do. For sub-solar metallicities, we find the opposite behaviour.

\begin{figure}
\centering
\includegraphics[width=1.0\linewidth]{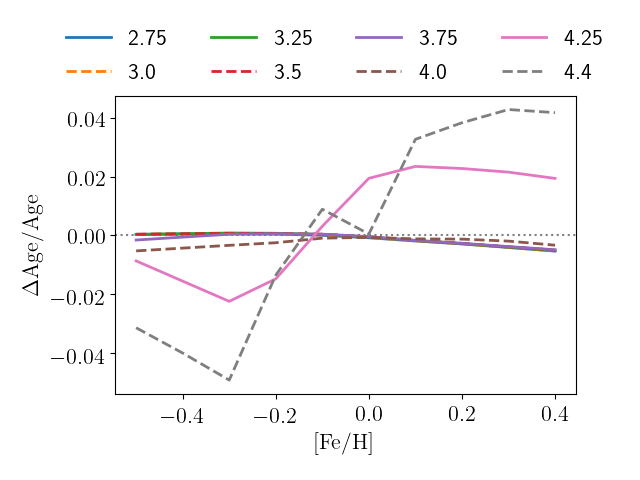}
\caption{Relative difference in age at fixed metallicities and surface gravities for stars with $1\,\mathrm{M}_\odot$. Positive residuals imply that our coupled model is older than its standard model counterpart. The surface gravity ($\log g$) is specified in the legend. To avoid complications arising from the red giant branch bump, we only include models for which $\log g >= 2.75$.}
\label{fig:Metal3}
\end{figure}

Figure~\ref{fig:Metal3} shows the difference in age between standard and coupled stellar models of $1\,\mathrm{M}_\odot$ stars with different surface gravities as a function metallicity. To ensure that we are comparing models with the exact same surface gravities, we interpolate in the global stellar parameters of the computed standard stellar models. We find that the largest absolute and relative age differences are obtained on the main sequence and close to the {\color{black}TO}, which implies that the use of standard stellar models affect isochrones and thus age estimates for clusters. The largest error on the main sequence is thus of the order of 4 per cent.

As regards Figs~\ref{fig:Metals1} and \ref{fig:Metal3}, we note that there is no difference in effective temperature or age on the main-sequence ($\log g = 4.4$) at solar metallicity by construction. At this point of its evolution, the corresponding $1\,\mathrm{M}_\odot$ star lies close to the present-day Sun, based on which the initial conditions of both grids were determined.


\section{Hare and hound exercise} \label{sec:hare}

In this section, {\color{black}we} perform an artifical asteroseismic analysis, in which we examine how well we can infer the global stellar properties of coupled models based on a grid of standard stellar models. For this purpose, we employ \textsc{aims} (cf. Section~\ref{sec:MCMC}). {\color{black}The aim of our hare and hound exercise is to evaluate the magnitude of the systematic biases that are introduced on the inferred parameters when using standard rather than coupled stellar models. While coupled models give a more physically realistic depiction of stars, it is yet to be demonstrated that coupled models also yield more accurate parameter estimates. We do not aim to settle this issue here. However, under the assumption that the properties of coupled models more closely represent those of real stars, our analysis can give us an idea of how well standard stellar models perform in actual asteroseismic analyses.}

As we have repeatedly addressed asteroseismic analyses of main-sequence stars in previous papers \citep{Joergensen2019, Mosumgaard2020}, we turn our attention to TO and RGB stars. This choice is important, since it affects how to construct a grid of stellar models for \textsc{aims} to interpolate in \citep[cf.][for a detailed discussion of this issue]{2019MNRAS.484..771R}.

We use models with the same input physics as used in Sections~\ref{sec:gasG1} and \ref{sec:global}. As regards the coupled stellar models, we consider a subsample from the grid, consisting of 68 models at solar metallicity with $\log g <4.2$. As regards the grid of standard stellar models, we include models with metallicity between $-0.5$ and $0.5\,$dex in steps of $0.1\,$dex. The mass of the standard models range between $0.7$ and $2.5\,\mathrm{M_\odot}$ in steps of $0.02\, M_\odot$ --- this is thus a different grid than the one used throughout Section~\ref{sec:global}.


\subsection{Likelihood} \label{sec:likeli}

{\color{black}We strive to recover a set of non-seismic properties ($L$, $\mathrm{[Fe/H]}$, and $T_\mathrm{eff}$) as well as the individual model frequencies of coupled stellar models using standard stellar models. For} each of the considered coupled stellar models, we computed adiabatic oscillation frequencies within the gas $\Gamma_1$ approximation using \textsc{adipls}. To account for the surface effect when using standard stellar models, we included the surface correction relation by \cite{Sonoi2015}: 
\begin{equation}
\frac{\delta \nu}{\nu_\mathrm{max}} =  \alpha \left(1-\frac{1}{1-(\nu_\mathrm{obs}/\nu_\mathrm{max})^{\beta}} \right). \label{eq:lorentz}
\end{equation}
Here, we let $\alpha$ be a free parameter but require it to be negative. We do so, in order to recover the notion that the adiabatic frequencies are assumed to overestimate the observed frequencies across the HR diagram analogously to the case of the present-day Sun \citep{Houdek2017, Houdek2019}.

In Eq.~(\ref{eq:lorentz}), we adopt 
\begin{equation}
\log \beta = - 3.86 \log T_\mathrm{eff} + 0.235 \log g + 14.2. \label{eq:logbeta}
\end{equation}
from \cite{Sonoi2015}. We note that Eqs~(\ref{eq:lorentz}) and (\ref{eq:logbeta}) have been calibrated based on the \textit{radial} modes ($\ell = 0$) of patched models. We thus limit ourselves to only include radial modes in the likelihood --- the results discussed in Section~\ref{sec:hhsun} constitute an exception to allow for a more direct comparison with \cite{2019MNRAS.484..771R}.

The reason for choosing the surface correction relation by \cite{Sonoi2015} is that it has been derived {\color{black}within the gas $\Gamma_1$ approximation} based on patched models rather than based on observations. {\color{black}For} the present-day Sun, this surface correction relation, therefore, recovers frequencies that closely resemble those obtained within the gas $\Gamma_1$ approximation in Section~\ref{sec:solar2} (cf. Fig.~\ref{fig:Sunfreqdiff}). So far as that we believe that the surface correction relation by \cite{Sonoi2015} \textit{generally} yields a good parameterization of the model frequencies of patched models, our coupled and standard stellar models, therefore, treat the surface effect consistently.
Based on an analysis of hundreds of patched models, for which the frequencies have been computed within the gas $\Gamma_1$ approximation, \cite{Joergensen2019} indeed demonstrate that a Lorentzian parameterization of the associated structural surface effect also performs well for giants and subgiants \citep[see also][]{Manchon2018}.

This being said, the surface correction relation by \cite{Sonoi2015} is based on only ten patched models. Moreover, these ten models predominantly correspond to main-sequence stars and subgiants with $T_\mathrm{eff} \geq 6000$; $\log g$ is only lower than 3.5 in two out of the ten samples. As a result this surface correction relation is subject to a selection bias, which might affect the inferred surface effect \citep{Joergensen2019, Joergensen2020}. However, the use of any other surface correction relation than that by \cite{Sonoi2015} would be problematic, since they do not recover the systematic frequency offset that haunts the gas $\Gamma_1$ approximation. 

To include the theoretical frequencies and the remaining artificial observables into the likelihood, we have ascribed artificial statistical errors to the properties of the coupled models. For all model properties, we assume the noise to be Gaussian. {This assumption is commonly used in the literature \citep[e.g.][]{Aguirre2015, Nsamba2018}. The likelihood ($\mathcal{L}$) thus takes the form
\begin{eqnarray}
\hspace{-1.5cm}\mathcal{L} &=& \left(2\upi\right)^{-N/2} \left|\mathbfss{C} \right|^{-N/2}
\times \nonumber\\
&& \hspace{-1.cm} \exp \left( -  \frac12 \sum_{i=1}^{N} \left[x_{\mathrm{1D},i}-x_{\mathrm{3D},i}\right]^T \mathbfss{C}^{-1} \left[x_{\mathrm{1D},i}-x_{\mathrm{3D},i}\right] \right).
{\label{eq:likeli}}
\end{eqnarray}
Here, the sum runs over all $N$ properties ($x_{\mathrm{3D},i}$) that we aim to recover from the coupled models, $x_{\mathrm{1D},i}$ denotes the corresponding model predictions from the standard stellar models, and $\mathbfss{C}$ denotes the co-variances. In this paper, we assume that the observed quantities, including the non-seismic constraints as well as the individual frequencies, are uncorrelated. Consequently, the expression inside the exponential in Eq.~\ref{eq:likeli} reduces to the expression for $-\chi^2/2$.}

Based on \cite{Lund2017}, we assume that a standard deviation of $0.1\,\mu$Hz can be achieved at $\nu_\mathrm{max}$ for $\nu_\mathrm{max}=2500\,\mu$Hz and that the same relative uncertainty can be achieved at $\nu_\mathrm{max}$ in general. For the remaining frequencies, we assume that the error increases quadratically with the frequency difference to $\nu_\mathrm{max}$, in order to mimic the decreasing amplitude of the modes:  
\begin{equation}
\sigma(\nu_{n,\ell=0}) = \frac{\nu_\mathrm{max}}{2500}\left(0.1 + 3.6\frac{(\nu_{n,\ell=0}-\nu_\mathrm{max})^2}{\nu_\mathrm{max}^2} \right).   
\end{equation}
Based on the assumption of a Gaussian envelope for the frequency amplitudes \citep[e.g.][]{Mosser2012,Rodrigues2017}, we only consider frequencies that deviate less than thrice the standard-deviation ($\sigma_\mathrm{env}$) of the Gaussian distribution from $\nu_\mathrm{max}$ \citep{Mosser2012}:
\begin{equation}
\sigma_\mathrm{env} = \frac{0.66\nu_\mathrm{max}^{0.88}}{2\sqrt{2\ln2}}. \label{eq:sigma}    
\end{equation}
As regards the remaining constraints, we set the uncertainty on $T_\mathrm{eff}$ and $\mathrm{[Fe/H]}$ to be $100\,$K and $0.1\,$dex, respectively. The uncertainty on the luminosity is set to be 3 per cent.


\subsubsection{The Sun} \label{sec:hhsun}

The chosen likelihood closely matches that used in analyses of real targets \citep[e.g.][]{Joergensen2020}. However, to further validate that the chosen likelihood leads to meaningful results, we ran the hare and hound exercise for the present-day Sun. For this purpose, we used the model frequencies of {\color{black}one of} the coupled solar \textsc{cl{\'e}s} model in Section~\ref{sec:solar} (case a in Table~\ref{tab:solcaltab}). However, since the Sun is not in our grid, we used the grid by \cite{2019MNRAS.484..771R}. We note that this grid is based on the solar mixture found by \cite{Grevesse1998} and that it is thereby not fully consistent with the assumptions that enter our solar calibration model.

\cite{2019MNRAS.484..771R} show that they are able to recover the global properties of the Sun based on observed solar frequencies. This was accomplished using radial and non-radial modes ($\ell = 0$, $\ell = 1$, and $\ell = 2$) in combination with non-seismic constraints. When following this approach, we find that we obtain the global solar parameters with equivalent accuracy based on the model frequencies of coupled stellar models. $M=(1.001\pm0.003)\,\mathrm{M_\odot}$ and $\mathrm{Age} = (4534 \pm 91) \,$Myr. Meanwhile, a lower accuracy is achieved when treating the Sun as a star on the lines described in Section~\ref{sec:likeli} using only radial modes ($\ell = 0$). {Here, we find that} $M=(0.971\pm 0.003)\,\mathrm{M_\odot}$ and $\mathrm{Age} =(4808\pm128) \,$Myr. The discrepancies in mass and age thus correspond to 3 and 5 per cent, respectively. The impaired accuracy of the fit is a natural consequence of the fact that we include less informative constraints into the likelihood.

We note that we can use the grid by \cite{2019MNRAS.484..771R} to fit the special case of the Sun because the grid is based on a solar calibration and because the coupled solar models in Section~\ref{sec:likeli} demonstrably recover the true solar structure with high accuracy. On the other hand, the input physics that underlies our grid of coupled stellar models differs significantly from the assumptions that enter the grid by \cite{2019MNRAS.484..771R}. This is especially true for the composition profiles --- that is, whether or not, say, atomic diffusion is included. Performing an hare and hound exercise based on other coupled models of main-sequence stars would thus lead to results that would be very hard to interpret. The analysis of the RGB stars presented below, on the other hand, does not suffer from this obstacle, since we use a grid of 1D standard stellar models that is fully consistent with the coupled stellar models, whose properties we seek to recover. The only difference lies in the treatment of the superadiabatic surface layers.


\subsection{Goodness of fit} \label{sec:gof}

In the following, we quote the model parameters of the maximum posteriori models, to which we refer as the best-fitting models. We furthermore quote the $1\,\sigma$ uncertainties based on the associated $68\,\%$ credibility intervals derived from the posterior probability distributions. By doing so, we assume that the posterior probability distribution is well approximated by a Gaussian.

To access the goodness of fit for the best-fitting standard model, we quote the reduced $\chi^2$-value:
\begin{equation}
\chi^2_\mathrm{red} = \frac{1}{N} \sum_{i=1}^{N} \frac{(x_{\mathrm{1D},i}-x_{\mathrm{3D},i})^2}{\sigma_{\mathrm{3D},i}^2}. \label{eq:chi2}   
\end{equation}
{\color{black}Best-fitting} standard models, for which $\chi^2_\mathrm{red}\approx 1$, reliably recover the required properties of the coupled stellar models. While values of $\chi^2_\mathrm{red} \ll 1$ point towards overfitting, values of $\chi^2_\mathrm{red} \gg 1$ reveal a poor fit. We thus discard all models, for which $\chi^2_\mathrm{red} > 4$. 

Note that we neither use the reduced $\chi^2$ to establish the best-fitting models nor to determine uncertainties. For this purpose, we use the mapped posterior probability density. Instead, we merely use $\chi^2$ to discard targets from the analysis.


\subsection{Results} \label{sec:results_harehound}

{\color{black}Based on the reduced $\chi^2$-values of the best-fitting standard stellar models, we discard 28 coupled models. This leaves us with a sample of 40 models. Since we raise caution about the inferred properties of coupled models near the TO in Section~\ref{sec:global}, it is worth mentioning that the majority of the 40 coupled models in our sample are giants and subgiants that lie far from the TO. Moreover, we note that the few coupled models that lie close to TO do not skew the sample or bias the qualitative and quantitative conclusions that are drawn in this section.}

\begin{figure}
\centering
\includegraphics[width=1.0\linewidth]{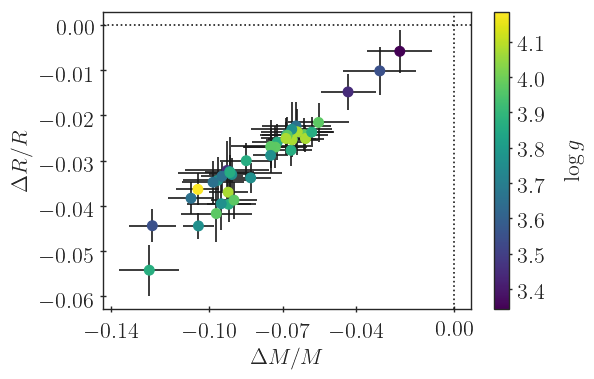}
\caption{Relative difference in mass and radius between the inferred values and the underlying parameters of the coupled models. Positive values imply that the mass and radius have been overestimated.  
}
\label{fig:HareHound}
\end{figure}

We find that the structural changes that are introduced by the improved boundary conditions are so large that we cannot accurately infer the global stellar properties of the underlying coupled stellar models from our grid of standard stellar models. We thus find that the mass and radius are consistently underestimated. We summarize these findings in Fig.~\ref{fig:HareHound}.

As can be seen from Fig.~\ref{fig:HareHound}, there is a clear correlation between the discrepancies in the inferred masses and radii. This reflects the fact that the best-fitting standard models approximately recover the mean densities of the underlying coupled models, since they are required to recover the individual mode frequencies.

{\color{black}In all 40 cases,} the surface correction relation by \cite{Sonoi2015} lowers the model frequencies. Moreover, the relative change in $\Delta \nu$ as a result of the surface correction is roughly constant {\color{black}as a function of} $\nu_\mathrm{max}$ and {\color{black}is} of the order of $10^{-2}$ for all models. The obtained behaviour of the inferred surface effect is thus similar to that obtained from asteroseismic analyses based on actual observations (e.g. \citealt{Rodrigues2017}, who use standard stellar models and scaling relations, and \citealt{Joergensen2020}, who use standard stellar models). 

\begin{figure}
\centering
\includegraphics[width=1.0\linewidth]{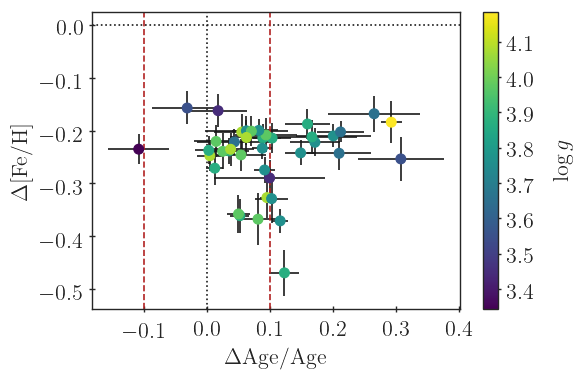}
\caption{Absolute difference in metallicity and fractional difference in age between the inferred values and the underlying parameters of the coupled models. Positive values imply that the metallicity and age have been overestimated. The dashed red line divides the sample into those models that recover the correct age within $10\,\%$ and those that do not. The plot contains 40 stars that passed our selection criteria.}
\label{fig:HareHound2}
\end{figure}

\begin{figure}
\centering
\includegraphics[width=1.0\linewidth]{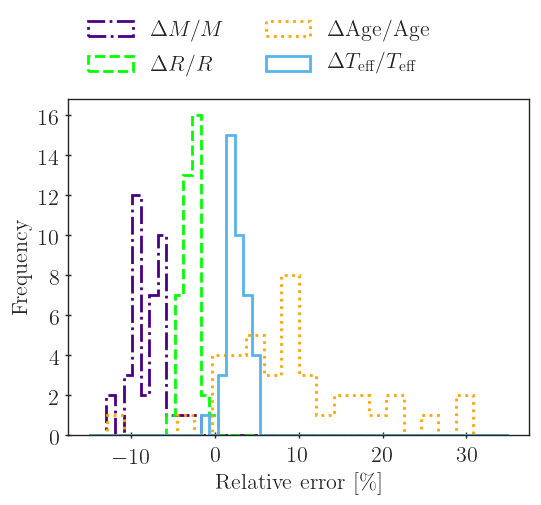}
\caption{Relative error in the inferred stellar parameters for the stellar mass, radius, age, and effective temperature based on the samples presented in Figs~\ref{fig:HareHound} and \ref{fig:HareHound2}.}
\label{fig:HareHoundTeff}
\end{figure}

{\color{black}We find that the ages of the coupled model are systematically overestimated and this often by more than $10$ per cent. On} average, the deviation in age is $8.8\pm1.8$ per cent. {\color{black}Furthermore, we find that} the metallicity is systematically underestimated --- on average by $(-0.23\pm0.04)\,$dex. {\color{black}We illustrate both of these findings in Fig.~\ref{fig:HareHound2}. The} effective temperature is systematically overestimated --- on average by $(150\pm 28)\,$K, i.e. $1.5\,\sigma$, {where $\sigma$ denotes the attributed observational error. We illustrate this latter statement in Fig.~\ref{fig:HareHoundTeff}.}

\cite{Tayar2017} have evaluated the shift in the mixing length parameter that is necessary to recover observation constraints on over 3000 red giants {\color{black}based on standard stellar models (see also Appendix~\ref{sec:tayar})}. Based on their analysis, \cite{Tayar2017} conclude that the omission of this correction can affect isochrone ages {\color{black}by} as much as a factor of two, even when considering target stars with near-solar metallicity. Whether or not the asserted variation in $\alpha_\mathrm{mlt}$ is indeed physical \citep{Salaris2018}, the results by \cite{Tayar2017} demonstrate the huge impact of the chosen input physics of stellar models on the derived parameter estimates. While the obtained errors in the stellar age in Fig.~\ref{fig:HareHound2} are high, they do thus not seem implausible. Stellar ages obtained from asteroseismic analyses based on standard stellar models may thus suffer from significant systematic errors. This is also reflected in the age uncertainties determined from methods such as those used by \cite{Bellinger2016} and \cite{Angelou2017} where the input physics is varied widely. Even when studying the present-day Sun as a star by attempting to recover the solar properties based on observations, changes in the input physics of the models can play a significant role as shown by e.g. \cite{2019MNRAS.484..771R}.

As demonstrated in Fig.~\ref{fig:Agediff}, the best-fitting standard stellar models of RGB stars should recover the correct age, if the models were to accurately predict the stellar masses, radii, and metallicities. It follows that the systematic errors in the inferred ages mirror the incorrectly deduced masses and radii, which in turn reflect the chosen constraints. {\color{black}We} thus repeated the analysis, substituting the constraint on the luminosity with a constraint on the stellar radius. In practice, such constraints are available from interferometric measurements or dynamical studies of binaries. Based {\color{black}on studies of eclipsing binaries by} \cite{White2013} and \cite{Gaulme2016}, we assumed that an error of two per cent is feasible. Doing so, however, we arrive at similar qualitative mismatches. This is due to the fact that the accuracy, with which the radius is recovered, is already of the order of two per cent without including the radius in the likelihood.

\begin{figure}
\centering
\includegraphics[width=1.0\linewidth]{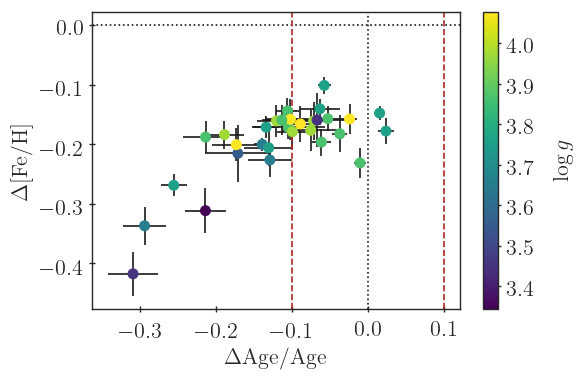}
\caption{As Fig.~\ref{fig:HareHound2} for a different likelihood: here, we have substituted the constraint on the stellar luminosity with constraints on both the stellar mass and radius. The plot contains 31 stars that passed our selection criteria. These are all also in the subsample in Figs~\ref{fig:HareHound}-\ref{fig:HareHound2}. Regarding the four outliers at low metallicity, we note that these are all early RGB stars with  $T_\mathrm{eff}\lesssim 4800\,$K with $\log g \lesssim 3.7\,$dex, for which the radius and mass estimates closely match the values of the underlying coupled models. The outliers are discussed further in the bulk text.
}
\label{fig:HareHound3}
\end{figure}

For some stars, dynamical studies provide robust observational constraints on the stellar mass. In the best-case scenario, the statistical errors are of the order of one per cent or lower \citep{Pourbaix2016,Gaulme2016}. Adopting these optimistic uncertainties on the mass, we repeated the analysis. This time the errors in the obtained masses and radii are recovered within $6\,\%$ and $3\,\%$, respectively, for all considered coupled models. Meanwhile, a large fraction of the RGB stars do still not recover the correct age within $10\,\%$, as shown in Fig.~\ref{fig:HareHound3}.

\begin{figure}
\centering
\includegraphics[width=1.0\linewidth]{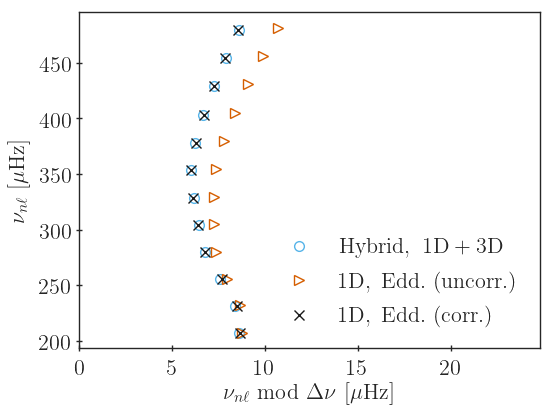}
\caption{Echelle diagram showing the adiabatic radial mode frequencies of a coupled model within the gas $\Gamma_1$ approximation and the uncorrected and corrected adiabatic frequencies of the associated
best-fitting standard model found by \textsc{aims} using the surface correction relation by \citet{Sonoi2015}. For the presented coupled model, $M = 0.91 \, \mathrm{M_\odot}$, $T_\mathrm{eff} = 4724\,$K, and $\log g = 3.45\,$dex. The model corresponds to the outlier, for which $\Delta \mathrm{[Fe/H]} = -0.41\pm0.04\,$dex in Fig.~\ref{fig:HareHound3}. {\color{black}As can be seen from the figure, the inferred surface effect behaves as expected. The same holds true for the other cases.}
}
\label{fig:echelle2}
\end{figure}

Comparing Figs~\ref{fig:HareHound2} and \ref{fig:HareHound3}, we note that the inferred stellar ages go from being too high in Fig.~\ref{fig:HareHound2} to being too low in Fig.~\ref{fig:HareHound3}. The inferred ages are hence sensitive to the change in the likelihood function. This implies that the individual frequencies do not dominate the likelihood. The observed age deviation can, therefore, not be explained by our treatment of the surface effect alone. {\color{black}The age deviations demonstrably reflect the} changes in the global stellar parameters that arise from the improved boundary conditions (cf. Section~\ref{sec:global}). This is not to say that the frequencies are not well-fitted. On the contrary, the echelle diagrams look as expected. We show this in Fig.~\ref{fig:echelle2}.

{\color{black}The} samples, for which we infer the largest age discrepancies in Fig.~\ref{fig:HareHound3}, also lead to the largest discrepancies in metallicity. Figure~\ref{fig:HareHound3} thus contains 4 outliers, for which the deviations in metallicity lie between -0.27 to $-0.41\,$dex, and for which the deviations in ages exceed that of the remaining 27 samples. There are, meanwhile, several (4) other models with similar $\log g$ ($<3.7\,$dex) yielding better age and metallicity estimates than the outliers. However, the higher accuracy in age and metallicity comes at the cost of lower accuracy in both the mass and radius. To fit the required stellar properties for the outliers, \textsc{aims} thus compensated for the impact of the different boundary conditions of standard and coupled stellar models on the global stellar properties by adjusting the stellar metallicity. Such offsets in metallicity are well-known to affect stellar age estimates and can, therefore, explain the associated large deviations in age \citep[e.g.][]{1994ApJS...95..107W,1999ASPC..192..283W}. However, even without taking these outliers into account, the mean deviation in age is still $5.8\pm1.5$ per cent. Moreover, ignoring the outliers, the offset in metallicity is roughly constant --- this holds true for both Figs~\ref{fig:HareHound2} and \ref{fig:HareHound3}.
The bias in age does hence not generally scale with the bias in metallicity.

We also repeated the analysis, including constraints on $T_\mathrm{eff}$, $\mathrm{[Fe/H]}$, and $\nu_\mathrm{max}$. Here, we set the error on $\nu_\mathrm{max}$ to one per cent. Once again, we reach the same qualitative conclusions regarding the offsets in mass, radius, metallicity, and age.

Independently of the constraints that enter our likelihood, we thus always end up with the same qualitative conclusion: the outer boundary conditions impact on the outcome of asteroseismic analyses through the resulting change in the global stellar parameters. However, the ability of standard stellar models to recover given properties of coupled stellar models are, of course, sensitive to these constraints. Indeed, it is well-known that the stellar parameters of asteroseismic analyses reflect the chosen likelihood \citep[e.g.][]{Aguirre2013,Basu2018,Nsamba2018}. 
To give the non-seismic constraints higher impact, one might shift to using global seismic constraints, such as the large frequency separation, rather than individual frequencies. Using this approach, \cite{JoergensenAngelou2019} are able to achieve mutually consistent parameter estimates for main-sequence stars based on coupled and standard stellar models --- the individual frequencies are meanwhile not perfectly recovered.
Alternatively, one might introduce more information from seismic constraints by drawing upon higher degree modes. This has been shown to be a successful strategy by \cite{Mosumgaard2020}, who recover very similar global properties for \textit{Kepler} stars using both coupled and standard stellar models. However, we also note that the modifications that are required for \textsc{aims} to correctly handle higher degree modes for more evolved stars lie beyond the scope of this paper. 

Note that the findings by \cite{JoergensenAngelou2019} and \cite{Mosumgaard2020} are consistent with our hare and hound exercise for the present-day Sun in Section~\ref{sec:hhsun}. Moreover, this statement does not contradict the conclusions drawn in this section, as we do not address main-sequence stars here.


\subsection{Discussion}

{\color{black}A direct comparison between actual observations and the individual model frequencies of coupled models is hampered by inaccurate nature of the gas $\Gamma_1$ approximation (cf. Figs~\ref{fig:Sunfreqdiff} and \ref{fig:GGvsRG}). The precision of the observed frequencies thus, by far, surpasses the accuracy of the model frequencies. For main-sequence stars, this issue can be avoided by circumventing the surface effect altogether. Rather than comparing observations to individual model frequencies, one might employ the frequency ratios proposed by \cite{Roxburgh2003}. These ratios are insensitive to the surface layers, as shown by \cite{Oti2005}. With this in mind, coupled stellar models can successfully be applied in analysis of real stars as shown by \cite{JoergensenAngelou2019} and \cite{Mosumgaard2020}. However, it is yet not settled, whether the use of frequency ratios is a safe and viable strategy beyond the main-sequence, due to the occurrence of mixed modes.}


\section{Conclusion}

In this paper, we discuss coupled stellar models that combine state-of-the-art one-dimensional standard stellar models with three-dimensional simulations of the outermost layers of convective envelopes. Our work is a continuation of a series of papers, in which we have established the robustness and versatility of our method \citep{Joergensen2017, Joergensen2018, Joergensen2019, JoergensenWeiss2019, JoergensenAngelou2019, Mosumgaard2020}. {Our results can be summarized as follows:

\begin{enumerate}
    \item \textbf{We show that the uncertainties on the global solar parameters that enter solar calibrations allow for shifts in the individual model frequencies of the order of $1-2\,\mu$Hz} (cf. Section~\ref{sec:solar}). With this finding, we are able to explain the differences between the model frequencies obtained from different {\color{black}coupled and patched} solar models that are presented in the literature \citep[cf.][]{Schou2020}. Moreover, we note that the remaining residuals between our coupled solar calibration models and observations lie below $2\,\mu$Hz at all frequencies. Coupled stellar models have thus reduced the surface effect to become comparable to the established error bars.
    \item \textbf{We demonstrate that the gas $\Gamma_1$ approximation generally performs well for low-mass main-sequence stars (cf. Section~\ref{sec:gasG1}).} {\color{black}For all stellar models within the explored region of the parameter space, the errors that are introduced by the gas $\Gamma_1$ approximation lie within $2.9\,\mu$Hz at $\nu_\mathrm{max}$.} The {\color{black}applicability} of this approximation beyond the case of the present-day Sun has previously not been {\color{black}validated}.
    \item \textbf{We find that the improved outer boundary layers of coupled models impact the predicted stellar properties across the HR diagram (cf. Section~\ref{sec:global})}. At fixed mass, age, and metallicity, the deviation between the effective temperatures of coupled stellar models and their standard stellar counterparts thus exceeds $80\,$K in some cases. The discrepancy in the stellar radius meanwhile ranges from a few per cent to $25$ per cent. Discrepancies in the mean density and large frequency separation reach 2 and 3 per cent, respectively.
    \item \textbf{In a hare and hound exercise, we demonstrate that the dissonance between standard and coupled stellar models affects the outcome of asteroseismic analyses (cf. Section~\ref{sec:hare}).} In this exercise, we attempt to recover the global stellar parameters of the coupled stellar models {\color{black}($3.4\lesssim \log g \lesssim 4.1$)} by drawing upon standard stellar models. We show that the inferred stellar properties deviate significantly from the ground truth. The deviation in the inferred stellar age thus often exceeds $10$ per cent, which corresponds to the desired accuracy of the PLATO space mission --- both for the core objectives of the mission and for the purpose of galactic archaeology \citep{Rauer2013, Miglio2017}.
\end{enumerate}
}

{\color{black}Although coupled stellar models give a more realistic depiction of the stellar surface layers than standard stellar models do, it is not settled whether coupled models also yield more accurate parameter estimates. However, our results demonstrate that the treatment of superadiabatic convection not only affects the model frequencies, but also alters the predicted global stellar parameters. 
In the light of the high-quality asteroseismic data from current and up-coming Earth-bound surveys and space missions, it is hence \textit{not} enough to address the surface effect when attempting to deal with the shortcomings of standard stellar models. One must also consider the impact of a simplified depiction of superadiabatic convection on stellar evolution. Our results thus strongly advocate a synergy of state-of-the-art stellar evolution codes and multi-dimensional simulations of magneto-hydrodynamics. Our coupled stellar models show a possible way towards achieving this synergy and thereby provide essential improvements towards the next generation of stellar models. As the next step in our exploration of coupled models, we will produce grids of coupled stellar models to be used in asteroseismic analyses of \textit{Kepler}, TESS and PLATO target stars.}


\section*{Acknowledgements}
{We acknowledge the useful feedback of our anonymous referee.} The research leading to this paper has received funding from the European Research Council (ERC grant agreement No.772293 for the project ASTEROCHRONOMETRY). Funding for the Stellar Astrophysics Centre is provided by The Danish National Research Foundation (Grant agreement No.~DNRF106). V.S.A. acknowledges support from the Independent Research Fund Denmark (Research grant 7027-00096B), and the Carlsberg foundation (grant agreement CF19-0649). JRM acknowledges support from the Carlsberg Foundation (grant agreement CF19-0649).

\section*{Data Availability Statement}

The data underlying this article will be shared on reasonable request to the corresponding author.




\bibliographystyle{mnras}
\bibliography{manual_refs,mendeley_export}



\appendix


\section{Overcoming interpolation errors} \label{sec:interr}

As discussed by \cite{Mosumgaard2020}, {\color{black}coupled stellar} models are subject to interpolation errors, due to the insufficient resolution of the underlying grids of 3D RHD simulations. In addition, we note that the stellar evolution codes rely on linear interpolation algorithms when computing the temperature and turbulent pressure profiles of coupled stellar models {\color{black}(cf. flowchart in Fig. 1 in \citealt{JoergensenWeiss2019})}. These algorithms recover the structure of the $\langle 3\mathrm{D} \rangle$-envelopes with lower accuracy than the algorithms presented by \cite{Joergensen2017}, since \cite{Joergensen2017} interpolate in the $(T_\mathrm{eff},\log g)$-plane by constructing piece-wise cubic, continuously differentiable surfaces \citep[see also Chapter 5 in ][]{ediss25003}. Both the insufficiently low resolution of the grid of 3D simulations and the linear interpolation ultimately lead to kinks in the evaluated stellar evolution tracks.

To overcome the errors that result from the linear interpolation, we have used a piece-wise cubic interpolation to produce a denser artificial grid of $\langle 3\mathrm{D} \rangle$-envelopes. This denser grid is now employed by our stellar structure and evolution code to construct stellar models using linear interpolation in this paper. 

\begin{figure}
\centering
\includegraphics[width=1.0\linewidth]{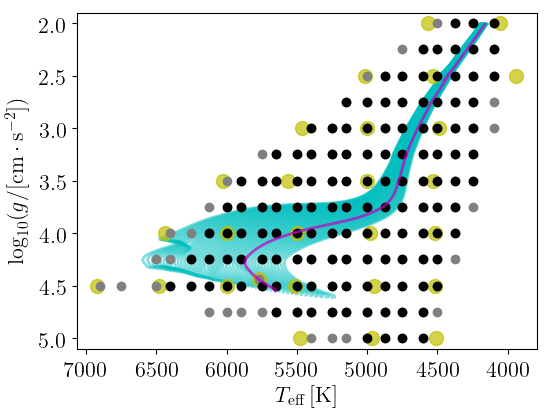}
\caption{Artificial grid of 3D RHD simulations. The black dots denote the simulations that are available at all metallicities. The grey dots show the extent of the grid in the $(T_\mathrm{eff},\log g)$-plane at solar metallicity. The purple track shows the evolution of a $1\,\mathrm{M}_\odot$ star at solar metallicity. The cyan lines show the evolution tracks of all models that enter the analysis in Section~\ref{sec:global} at solar metallicity. These tracks span masses between 0.88 and $1.32\,\mathrm{M}_\odot$. The location of the original 3D simulations in the Stagger grid are marked with yellow circles.
}
\label{fig:tracks}
\end{figure}

Our artificial grid of $\langle \mathrm{3D} \rangle$-envelopes is shown in Fig~\ref{fig:tracks} together with stellar evolution tracks. We find that the use of the artificial denser grid greatly improves the stellar evolution tracks, especially on the red giant branch (RGB). The un-physical kinks on the RGB have been reduced to such an extent that we deem an analysis of red giants with our coupled models both meaningful and feasible. {\color{black}Although} the un-physical kinks have partly disappeared, the tracks still show small irregularities. The issue of the kinks has thus not been entirely solved by the use of the artificial grid, and
an extension of the Stagger grid is still much desired to further improve the interpolation \citep[see also ][]{Joergensen2017,Joergensen2019}. The Stagger grid has a spacing in $\log g$ and $T_\mathrm{eff}$ of $0.5\,$dex and 500$\,$K, respectively. Especially on the RGB, additional 3D simulations at intermediate values of $\log g$ and $T_\mathrm{eff}$ are required.

The use of such artificial grids also allows for a straight-forward rudimentary interpolation in metallicity at every time-step of the evolution and at a low computational cost. {\color{black}One} can compute artificial grids with any metallicity, including intermediate metallicities that do not exist in the underlying grid of 3D simulations, by employing the interpolation scheme presented by \cite{Joergensen2019}. At every time step, the stellar evolution code can then select that artificial grid of models that matches the current metallicity of the convective envelope, achieving a discrete resolution in $\mathrm{[Fe/H]}$. We have introduced this scheme in both \textsc{garstec} and the \textsc{cl{\'e}s} stellar evolution code.

{\color{black}Since} the stellar evolution code chooses that artificial grid, whose metallicity most closely matches the composition {\color{black}of the convective envelope} at every time-step, {\color{black}the code} can to some extent adapt to composition changes that arise from {\color{black}atomic} diffusion. In this manner, we address changes in metallicity from one time-step to the next. In this paper, we have computed artificial grids with a resolution in $\mathrm{[Fe/H]}$ of $0.1\,$dex, but with the exception of Sections~\ref{sec:solar} we do not include atomic diffusion. 

In connection with the chemical composition of the models, we note that the Stagger grid does not contain models with different helium contents at fixed metallicity. While one could partly account for this by introducing an offset between the surface gravity of the appended $\langle \mathrm{3D} \rangle$-envelope and the interior \citep{2013ApJ...778..117T}, this issue limits the degree, to which coupled models can in practice account for variations in the chemical profile. We also note that the helium abundance in the 3D simulations decreases with increasing metallicity. This means that the composition of the $\langle \mathrm{3D} \rangle$-envelopes is slightly at odds with the chemical evolution of the galaxy when addressing stars with a composition that deviates from that of the present-day Sun. These considerations call for further extensions of the Stagger grid to increase its versatility in connection with stellar evolution calculations.

Like for the interpolation in the $(T_\mathrm{eff},\log g)$-plane, additional 3D simulations at intermediate values are still needed to improve the interpolation in metallicity. For a discussion on how well the interpolation in metallicity performs across the Stagger grid, we refer the reader to \citep{Joergensen2019}. 

Rather than introducing an artificial grid of 3D simulations, one might as well include higher-order interpolation schemes directly into the stellar evolution code. Although this might come at a higher computational cost, it is doubtlessly a viable solution. A thorough exploration of different higher-order interpolation techniques is, however, beyond the scope of this paper. Here, we have settled for the mentioned approach based on artificial grids. From this exercise, we can already conclude that the coupling scheme by \cite{Joergensen2018} and \cite{JoergensenWeiss2019} in tandem with the interpolation scheme by \cite{Joergensen2017} performs very well across the entire parameter space. The accuracy of the interpolation scheme by \cite{Joergensen2017} is, meanwhile, affected by the order of the interpolation that underlies its implementation as well as the employed triangulation \citep[cf.][]{Mosumgaard2020}.


\section{Varying the mixing length parameter} \label{sec:tayar}

In this paper, we assume that the mixing length parameter is constant throughout the computed stellar evolution. However, different studies demonstrate the need to vary $\alpha_\textsc{mlt}$ across the HR diagram when attempting to encapsulate the correct properties of stellar structures and their evolution using standard stellar models. Indeed, by varying $\alpha_\textsc{mlt}$, one might partly counteract the impact of the improper outer boundary layers of standard stellar models. This being said, we note that it is still not settled how to vary the mixing length parameter across the HR diagram and throughout the stellar interior \citep[e.g.][]{Schlattl1997,Ludwig1997,Ludwig1999,Trampedach2014b,Magic2015,Tayar2017,Sonoi2019,2020MNRAS.493.4987A}.

\cite{Tayar2017} investigate over 3000 red giants with $\log g$ between approximately 1.5 and 3.5, in order to evaluate what changes are needed in $\alpha_\textsc{mlt}$ for standard stellar models to recover the observational constraints from both APOGEE and \textit{Kepler}. They conclude that $\alpha_\textsc{mlt}$ is sensitive to $\mathrm{[Fe/H]}$. Concretely, they state that a change in the metallicity of $1.0\,$dex requires a change in $\alpha_\textsc{mlt}$ of 0.2 on the RGB.

\begin{figure}
\centering
\includegraphics[width=1.0\linewidth]{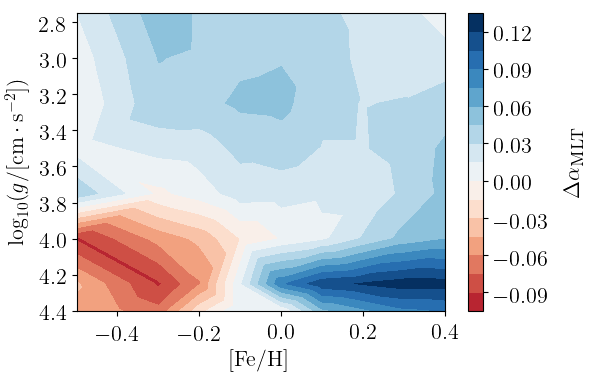}
\caption{Variation in the mixing length parameter that is required for the standard stellar models to recover the effective temperature at different $\log g$ for a $1.0\,\mathrm{M}_\odot$ {coupled model}. We only include models, for which $\log g \geq 2.75$ to avoid ambiguities that arise from the red giant branch bump.}
\label{fig:Alpha1}
\end{figure}

To address the statement by \cite{Tayar2017} and to contribute to the discussion regarding the metallicity dependence of the mixing length parameter, we have computed a set of standard stellar models, varying the $\alpha_\textsc{mlt}$ around the solar calibrated value. Mirroring the approach by \cite{Tayar2017}, we have then selected that value of $\alpha_\textsc{mlt}$ that recovers the value $T_\mathrm{eff}$ of coupled stellar models for different values of $\log g$. In all cases, we have compared coupled and standard stellar models with the same mass. For simplicity, we have fixed the mass to $1.0\,\mathrm{M_\odot}$. We varied $\Delta \alpha_\textsc{mlt}$ between $-0.18$ and $+0.18$ in steps of 0.02. The results are summarized in Fig.~\ref{fig:Alpha1}.

As can be seen from the figure, the variation of $\alpha_\textsc{mlt}$ on the main sequence is consistent with the variation found by \cite{Tayar2017}. However, for red giants, we do not find the same behaviour. Indeed, on the RGB, we only observe a limited variation in $\Delta \alpha_\textsc{mlt}$ with metallicity. This makes sense in the light of the results presented in Section~\ref{sec:acrossFeH}, where we find that the discrepancy in $T_\mathrm{eff}$ varies less on the RGB than on the main-sequence as a function of metallicity. In Section~\ref{sec:acrossFeH}, we, furthermore, find that the discrepancy in $T_\mathrm{eff}$ does not change sign on the RGB, i.e. the standard stellar models are consistently ($\leq 40  \,$K) too cold.

Our results are consistent with those of \cite{Salaris2018}, who re-analysed the stars addressed by \cite{Tayar2017} and found that \cite{Tayar2017} had not accounted for alpha enhancement. When only considering stars with scaled solar metal mixture, i.e. low $\alpha$-enhancement, \cite{Salaris2018} do not recover the metal dependence of the calibrated mixing length parameter. Meanwhile, \cite{Salaris2018} note that the metal dependence of $\alpha_\textsc{mlt}$ found by \cite{Tayar2017} is reintroduced when $\alpha$-enhanced stars are included.

\begin{figure}
\centering
\includegraphics[width=1.0\linewidth]{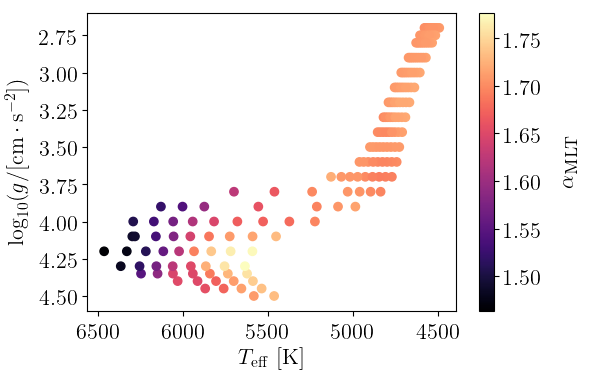}
\caption{Kiel diagram, showing the variation in $\alpha_\textsc{mlt}$ that is needed for the standard stellar models to recover the same position in the Kiel diagram as their coupled stellar model counterparts. For all models, $\mathrm{[Fe/H]=0.0}$. We compare standard and coupled stellar models with the same mass. Two outliers on the subgiant branch with $\log g=3.9$ as well as one outlier at $\log g=3.8$ were excluded from the plot. The mixing length parameter obtained for standard stellar model from a solar calibration with the same input physics is 1.67.}
\label{fig:Alpha2}
\end{figure}

\cite{Trampedach2014b}, \cite{Magic2015}, and \cite{Sonoi2019} have investigated the variation of $\alpha_\textsc{mlt}$ across the Kiel diagram based on 3D RHD simulations. To investigate the variation of $\alpha_\textsc{mlt}$ across the Kiel diagram and to compare to \cite{Trampedach2014b}, \cite{Magic2015}, and \cite{Sonoi2019}, we follow a similar approach to that used for the analysis in connection with Fig.~\ref{fig:Alpha1}. We thus compute a set of standard stellar models, for which we vary $\alpha_\textsc{mlt}$ around the solar calibrated value. This time, we vary the stellar mass but keep the metallicity fixed to the solar value. For the standard stellar models, we thus vary $\Delta \alpha_\textsc{mlt}$ between $-0.30$ and $+0.30$ in steps of 0.02 and vary the mass from $0.90$ to $1.3\,\mathrm{M_\odot}$ in steps of $0.05$. Again, we compare standard and coupled stellar models with the same mass. Following this procedure, our analysis of the variation in the $(T_\mathrm{eff},\log g)$-plane matches the approaches that underlie the investigation of the metallicity dependence presented in Fig.~\ref{fig:Alpha1}. The results are summarized in Fig.~\ref{fig:Alpha2}.

We find the same overall qualitative trends {\color{black}as} \cite{Trampedach2014b}, \cite{Magic2015}, and \cite{Sonoi2019} \citep[cf. Fig.~4 and Fig.~2 and Table~A.1 in][respectively]{Trampedach2014b, Magic2015, Sonoi2019}. Thus, we find that lower values of $\alpha_\textsc{mlt}$ are generally needed at higher effective temperatures when considering stars at the same $\log g$. To quantify this statement, we compared our results for the variation of $\alpha_\textsc{mlt}$ with the corresponding results obtained by \cite{Magic2015} and \cite{Sonoi2019}. To facilitate a meaningful comparison, we limited ourselves to those cases, for which the underlying 3D simulations used by \cite{Magic2015} and \cite{Sonoi2019} lie within the region covered in Fig.~\ref{fig:Alpha2}. There are ten such cases, excluding the present-day Sun. In all ten cases, we find that we infer the same values for $\alpha_\textsc{mlt}$ as \cite{Magic2015} and \cite{Sonoi2019} do within 0.04. Furthermore, in all ten cases, we find the inferred shift in $\alpha_\textsc{mlt}$ to have the same sign as the corresponding shifts found by \cite{Magic2015} and \cite{Sonoi2019}. The fact that our results mirror those found by other authors that use different methods further underlines the validity and flexibility of our coupling scheme.


\bsp    
\label{lastpage}
\end{document}